\documentclass[12pt]{article}

\usepackage[superscript,nomove]{cite}
\usepackage{graphicx}
\usepackage{upgreek}
\usepackage{amsmath}
\usepackage{hyperref}
\hypersetup{
    colorlinks=true,
    citecolor=magenta,
    linkcolor=magenta,
    filecolor=magenta,      
    urlcolor=cyan,
}

\makeatletter
\renewcommand\@biblabel[1]{#1.}

\usepackage{times}

\newcommand{\eps}{\varepsilon}
\renewcommand{\figurename}{Fig.}

\newcommand{\mytitle}{Asymmetry underlies stability in power grids}
\title{\mytitle}

\author{
\hspace{-6mm}Ferenc~Molnar,$^{1}$
Takashi Nishikawa,$^{1,2,\ast}$
Adilson E. Motter$^{1,2}$\\
\\
\hspace{-6mm}\normalsize{$^{1}$Department of Physics and Astronomy, Northwestern University, Evanston, IL 60208, USA}\\
\hspace{-6mm}\normalsize{$^{2}$Northwestern Institute on Complex Systems, Northwestern University, Evanston, IL 60208, USA}\\
\hspace{-6mm}\normalsize{$^{\ast}$email: t-nishikawa@northwestern.edu}%
}

\date{}

\begin{document} 

\maketitle 

\baselineskip17.9pt

\section*{Abstract}

Behavioral homogeneity is often critical for the functioning of network systems of interacting entities.  In power grids, whose stable operation requires generator frequencies to be 
syn\-chro\-nized---and thus homogeneous---across the network, previous work suggests that the stability of 
synchronous states
can be improved by making the generators homogeneous.  Here, we show that a substantial additional improvement is possible by instead making the generators suitably heterogeneous.  We develop a general method for attributing this counterintuitive effect to \textit{converse symmetry breaking}, a recently established phenomenon in which the system must be asymmetric to maintain a stable symmetric state.  These findings constitute the first demonstration of converse symmetry breaking in real-world systems, and our method promises to enable identification of this phenomenon in other networks whose functions 
rely on
behavioral homogeneity.

\vfill\begin{center}

Nature Communications \textbf{12}, 1457 (2021)\\[1mm]
{\small The published version is available online at:\\[-1mm] \url{https://doi.org/10.1038/s41467-021-21290-5}}
\end{center}

\clearpage

\section*{Introduction}
\addcontentsline{toc}{section}{Introduction}

In an alternating current power grid, the generators provide electrical power that oscillates in time as sinusoidal waves.
Because these waves are superimposed before reaching the consumers, they need to be synchronized to the same frequency; otherwise, time-dependent cancellation between these waves would cause the delivered power to fluctuate, 
which can lead
to equipment malfunction and damage \cite{Machowski:2020}.  Maintaining frequency synchronization is challenging because the system is complex in various ways, with every generator responding differently to the 
continual
influence of disturbances and varying conditions~\cite{Backhaus:2013}.  Adding to the challenge is the increase in perturbations resulting from the ongoing integration of energy from intermittent sources \cite{Schafer:18}, the emergence of grid-connected microgrids \cite{Olivares:2014}, and the expansion of an increasingly open electricity market \cite{Griffin:2009}.  Furthermore, the inherent heterogeneities in the parameters of system components and in the structure of the interaction network are perceived as obstacles to achieving synchronization. 
Consistent with the view that heterogeneities may generally inhibit frequency homogeneity, an earlier study showed that homogenizing the (otherwise heterogeneous) values of generator parameters can lead to stronger stability of 
synchronous states than in the original system
\cite{Mot:13}.  An outstanding question remains, however, as to whether there is a heterogeneous parameter assignment (different from the nominal one) that would enable even stronger stability for 
synchronous states
than the best homogeneous parameter assignment.  Though motivated by its significance for power grids, this question is broadly relevant for improving the stability of homogeneous dynamics in complex network systems in general, including consensus dynamics in networks of human or robotic agents \cite{Judd:2010,Ren:2010}, coordinated spiking of neurons in the brain \cite{Axmacher2006,Penn:2016}, and synchronization in communication networks \cite{Bregni:2002,Wu:2010}.

\begin{figure}
\centering
\includegraphics[width=0.8\linewidth]{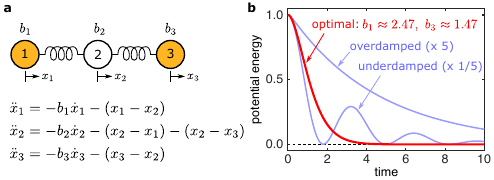}
\caption{Stabilizing effect of heterogeneity in a mass-spring system.  \textbf{a} System consisting of a linear chain of three unit masses connected by two identical springs.  The masses are constrained to move horizontally, and their dynamics are governed by the equation shown, where $x_i$ is the displacement of mass $i$ relative to its equilibrium and $b_i$ is its damping coefficient.  \textbf{b} Total potential energy of the springs vs.\ time for three different damping scenarios.  
The optimal damping (red), corresponding to the fastest energy decay, is achieved for $b_1 \approx 2.47$, $b_2 \approx 3.17$,  $b_3 \approx 1.47$ (or equivalently, $b_1 \approx 1.47$, $b_2 \approx 3.17$, $b_3 \approx 2.47$), despite the fact that masses $1$ and $3$ are otherwise identical and identically coupled.
Overdamping leads to a slower monotonic decay, while underdamping results in a slower oscillatory decay, as shown in blue by varying $b_1$ and $b_3$ by a factor of $5$.  In all cases, the initial conditions are $(x_1,x_2,x_3)=(1,0,-1)$ and $(\dot{x}_1,\dot{x}_2,\dot{x}_3)=(0,0,0)$.}
\label{fig_mass_spring_example}
\end{figure}

To gain insights into the potential role of heterogeneity in enhancing stability, it is instructive to first consider the case of damped harmonic oscillators.  For a single oscillator, the optimal stability corresponds to the fastest convergence to the stable equilibrium and is achieved when the oscillator is critically damped: underdamping would lead to lingering oscillations around the equilibrium, and overdamping would lead to slowed convergence due to excess dragging.  This optimization is exploited in door closers (devices that passively close doors in a controlled manner), which are designed to be critically damped for the door to close fast without slamming.  When multiple damped oscillators are coupled, the damping giving rise to optimal stability will be influenced by the network interactions.  More important, we can show that the optimal stability in such a network requires different oscillators to have different damping (even when their other parameters are all identical and they are positioned identically in the network), as illustrated in Fig.~\ref{fig_mass_spring_example}.

In this paper, we first demonstrate that an analogous effect occurs in power-grid networks: heterogeneity in generator parameters can robustly enhance both the linear and the nonlinear stability of 
synchronous states
in power grids from North America and Europe.  Since these systems have heterogeneity in the network structure in addition to the tunable generator parameters, 
one possibility is that the effect arises entirely from compensation: stability reduction due to one heterogeneity is compensated by another heterogeneity, leading to a stability enhancement when the latter heterogeneity is added.
An alternative, which we validate
here, involves the recently established phenomenon of \textit{converse symmetry breaking} (CSB) \cite{TakashiAIS}, in which the stability of a symmetric state requires the system's symmetry to be broken.
Due to its counterintuitive nature, this phenomenon had not been recognized until it was recently predicted and experimentally confirmed \cite{TakashiAIS,Nishikawa:2020} for synchronization in oscillator networks (a class of network dynamics widely studied in the literature \cite{Kiss:2007,Arenas:2008,Matheny:2019}).  Despite its conceptual generality and potential to underlie symmetric states of many systems, this phenomenon has not yet been observed outside laboratory settings.
The symmetry relevant here is node-permutation symmetry, since in a synchronized state the states of different nodes are equal and can be permuted without altering the state of the system.  For power grids, CSB would translate to a stability enhancement mechanism in which maintaining the stability of 
synchronous
(and thus symmetric) states requires the generator parameters to be heterogeneous (thus making the system asymmetric).  By systematically removing all the other system heterogeneities and isolating the effect of the generator heterogeneity, we establish that CSB is responsible for a significant portion of the stability improvement observed in the power grids we consider.  This offers insights into mechanisms underlying the parameter heterogeneity that arises when the generators are tuned to damp oscillations \cite{Rogers:2012,Obaid:2017} (e.g., by adjusting devices called power system stabilizers).  Our results are of particular relevance given that CSB has thus far not been observed in any real-world system outside controlled laboratory conditions, let alone power-grid networks.

\section*{Results}
\addcontentsline{toc}{section}{Results}

\paragraph{Power-grid dynamics and stability.}
\addcontentsline{toc}{subsection}{Power-grid dynamics and stability}
To describe the dynamics of $n$ generators in a power-grid network, we represent each generator node as a constant voltage source behind a reactance (the so-called classical model) and their interactions through intermediate non-generator nodes as effective impedances (a process known as Kron reduction) \cite{And:03}.
We assume that the system is operating near a synchronous state in which the voltage frequencies of the $n$ generators are all equal to a constant reference frequency $\omega_\mathrm{s}$, and we examine whether the homogeneous state is stable against dynamical perturbations.  Such perturbations, whether they are small or large, may come for instance from sudden changes in generation and/or demand due to shifting weather condition at wind or solar farms, variations in power consumption, switching on/off connections to microgrids, etc.
The short-term dynamics (of the order of one second or less) are then governed by the so-called swing equation \cite{And:03,Nish:15}:
\begin{equation}\label{eq-en}
\ddot{\delta_i} + \beta_i \dot{\delta_i} = a_i - \sum_{k \neq i} c_{ik} \sin \left(\delta_i - \delta_k - \gamma_{ik} \right),
\end{equation}
where $\delta_i$ is the phase angle variable for generator $i$ (representing the generator's internal electrical angle, relative to a reference frame rotating at the reference frequency $\omega_\mathrm{s}$); $\beta_i \equiv D_i/(2H_i)$ is an effective damping parameter (corresponding to $b_i$ in the mass-spring system of Fig.~\ref{fig_mass_spring_example}), with constant $D_i$ capturing both mechanical and electrical damping and constant $H_i$ representing the generator's inertia; $a_i$ is a parameter representing the net power driving the generator (i.e., the mechanical power provided to the generator, minus the power demanded by the network, including loss due to damping); and $c_{ik}$ and $\gamma_{ik}$ are respectively the coupling strength and phase shift characterizing the electrical interactions between the generators.
The parameters in Eq.~\eqref{eq-en} for a given system are determined by computing the active and reactive power flows between network nodes from system data and using them to calculate the complex-valued effective interaction (and thus its magnitude $c_{ik}$ and angle $\gamma_{ik}$) between every pair of generators.
In real power grids, stable system operation is ensured by a hierarchy of controllers that adjust generator power outputs and thus the parameters in Eq.~\eqref{eq-en}.  Here, however, these parameters can be regarded as constants, since the lowest level of control (known as the primary control) is modeled as a damping-like effect captured by the $\beta_i$ term in Eq.~\eqref{eq-en}, while the upper-level controls (known as the secondary and tertiary controls) act on time scales much longer than that of the short-term generator dynamics described by the model.  In addition, fluctuations in power generation and demand on the time scales of minutes or longer (which can come, e.g., from renewable energy sources) do not affect the short-term dynamics.
Equation~\eqref{eq-en} has recently been studied extensively in the network dynamics community \cite{pg3,Men:14,Schafer:18,Mot:13,Schafer:2018b,Yang:2017}.

We first analyze the stability of the synchronous state against small perturbations.  The synchronous state corresponds to a fixed point of Eq.~\eqref{eq-en} given by $\delta_i = \delta_i^*$ and $\dot{\delta}_i = 0$, which represents frequency synchronization because $\dot{\delta}_i$ is the frequency relative to the reference $\omega_\mathrm{s}$.
The Jacobian matrix of Eq.~\eqref{eq-en} at this point can be written as
\begin{equation}\label{eqn:J}
\mathbf{J} = \begin{pmatrix}
\,\,\,\,\mathbf{O} & \,\,\,\,\mathbf{I}\\
-\mathbf{P} & -\mathbf{B}
\end{pmatrix},
\end{equation}
where $\mathbf{O}$ and $\mathbf{I}$ denote the $n \times n$ null and identity matrices, respectively; $\mathbf{P} = (P_{ik})$ is the $n \times n$ matrix defined by
\begin{equation}\label{eqn:P}
P_{ik} = \begin{cases}
- c_{ik} \cos (\delta_i^* - \delta_k^* - \gamma_{ik}), & i \neq k, \\
-\sum_{k' \neq i} P_{ik'}, & i = k,
\end{cases}\vspace{3mm}
\end{equation}
which expresses
the effective interactions between the generators; and $\mathbf{B}$ is the $n \times n$ diagonal matrix with $\beta_i$ as its diagonal elements.  We note that, while the form of the Jacobian matrix for coupled damped harmonic oscillators is the same as in Eq.~\eqref{eqn:J}, power grids are different in that they can have $\mathbf{P}\neq\mathbf{P}^T$ because $c_{ik} \neq c_{ki}$ in general and because $\gamma_{ik}$ appears in Eq.~\eqref{eqn:P}.  The stability under noiseless conditions is determined by the Lyapunov exponent defined as $\lambda^\mathrm{max} \equiv \max_{i \ge 2} \mathrm{Re} (\lambda_i)$, where $\lambda_i$ are the eigenvalues of $\mathbf{J}$.  The identically zero eigenvalue, which comes from the zero row-sum property of $\mathbf{P}$ and is denoted here by $\lambda_1$, is excluded because it is associated with the invariance of the equation under uniform shift of phases.  If $\lambda^\mathrm{max} < 0$, then the 
synchronous state
is asymptotically stable, and smaller $\lambda^\mathrm{max}$ implies stronger stability. (This is known as small-signal stability analysis in power system engineering.)
Since real power-grid dynamics are noisy due to power generation/demand fluctuations and various other disturbances occurring on short time scales, $\lambda^\mathrm{max}$ needs to be sufficiently negative to keep the system close to the 
synchronous state.
Indeed, a previous study \cite{Nishikawa:2020} showed that, for broad classes of noise dynamics, there is a (negative) threshold 
value of $\lambda^\mathrm{max}$
for such stability: the system is stable if and only if $\lambda^\mathrm{max}$ is below the threshold.  
This stability threshold depends on the noise intensity level.  For impulse-like disturbances, the intensity level corresponds to the maximum deviation of $\delta_i$ that can be induced by a single disturbance, such as a sudden loss of a generator or a spike in power demand.  For continual disturbances, the intensity level can be quantified by the variances of the fluctuating power generation and demand, which can be modeled by adding a randomly varying term to the parameter $a_i$.  Since the stability threshold is generally lower for higher noise levels, the lower the value of $\lambda^\mathrm{max}$ for a given power grid, the more intense disturbances and fluctuations the system can endure without losing stability.
Incidentally, the optimal damping in the mass-spring system of Fig.~\ref{fig_mass_spring_example} is given precisely by minimizing $\lambda^{\max}$ for that system.

\paragraph{Enhancing stability with generator heterogeneity.}
\addcontentsline{toc}{subsection}{Enhancing stability with generator heterogeneity}
We now study $\lambda^\mathrm{max} = \lambda^\mathrm{max}(\boldsymbol{\upbeta})$ as a function of $\boldsymbol{\upbeta} \equiv (\beta_1,\ldots,\beta_n)$ for a selection of power grids whose dynamics can be described by Eq.~\eqref{eq-en} with the parameter values based on data.  Using the same model, it was previously shown \cite{Mot:13} that, under the constraint that all $\beta_i$'s have the same value, $\lambda^\mathrm{max}$ is minimized when $\boldsymbol{\upbeta} = \boldsymbol{\upbeta}_{=}$, where $\boldsymbol{\upbeta}_{=} \equiv (\beta_{=},\ldots,\beta_{=})$ and $\beta_{=} \equiv 2 \sqrt{\alpha_2}$, with $\alpha_2$ denoting the smallest nonidentically zero eigenvalue of matrix $\mathbf{P}$.  The eigenvalue $\alpha_2$ is associated with the least stable eigenmode, and we assume that it is real and positive (as confirmed in all systems we consider).  It was further shown that, at this homogeneous optimal point $\boldsymbol{\upbeta}_{=}$, the function $\lambda^\mathrm{max}(\boldsymbol{\upbeta})$ is non-differentiable (which precludes the use of a standard derivative test), but its one-sided derivative along any given straight-line direction is positive, i.e., the directional derivative $D_{\boldsymbol{\upbeta}'} \lambda^{\max} (\boldsymbol{\upbeta}_{=})$ is positive in the direction of any $n$-dimensional vector $\boldsymbol{\upbeta}'$.  Thus, moving away from $\boldsymbol{\upbeta}_{=}$ along any straight line would necessarily increase $\lambda^\mathrm{max}$ from the local minimum value $\lambda^{\max}(\boldsymbol{\upbeta}_{=}) = -\sqrt{\alpha_2}$ and hence only reduce the stability of the 
synchronous state.

\begin{figure}
\centering
\includegraphics[width=0.8\linewidth]{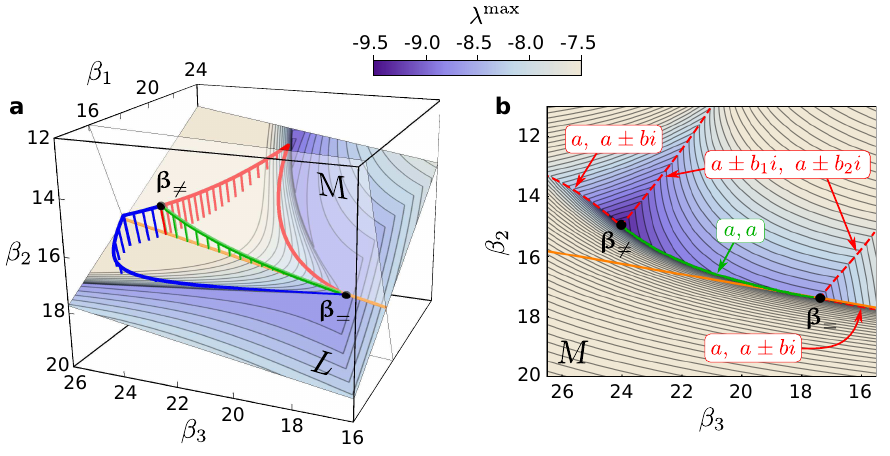}
\caption{\textbf{Enhancing the stability of the 3-generator system along curved paths.}  \textbf{a}~Examples of curved paths (red, green, and blue) in the $\boldsymbol{\upbeta}$-space from $\boldsymbol{\upbeta}_{=}$ to $\boldsymbol{\upbeta}_{\neq}$ along which the Lyapunov exponent $\lambda^{\max}$ decreases monotonically.  The numerically generated curves confirm our theoretical prediction that these paths are all tangent to the plane $L$ at the point $\boldsymbol{\upbeta}_{=}$.  The droplines indicate that these curves are outside $L$.  The contour levels of $\lambda^{\max}$ are shown on $L$.  Also shown is the plane $M$, which is defined as the plane perpendicular to $L$ that contains $\boldsymbol{\upbeta}_{=}$ and $\boldsymbol{\upbeta}_{\neq}$.  \textbf{b}~Contour levels of $\lambda^{\max}$ on the plane $M$, which contains the green path from \textbf{a}.  The orange lines in \textbf{a} and \textbf{b} indicate the intersection between planes $L$ and $M$.
The red dashed curves (and the green path) trace cusp surfaces associated with degeneracies of the real parts of the eigenvalues of $\mathbf{J}$ that determine $\lambda^{\max}$.  Three types of degeneracy are indicated: a real eigenvalue equal to the real parts of a pair of complex conjugate eigenvalues ($a$, $a \pm bi$), two different complex conjugate eigenvalues with equal real parts ($a \pm b_1i$, $a \pm b_2i$), and two equal real eigenvalues ($a$, $a$).
For details on the system, see Methods.}
\label{fig4}
\end{figure}

Despite the apparent impossibility of improving on $\lambda^{\max}(\boldsymbol{\upbeta}_{=})$ locally, we first show that there can be curved paths starting at $\boldsymbol{\upbeta}_{=}$ along which $\lambda^{\max}$ can be further minimized with heterogeneous $\beta_i$.  Indeed, Fig.~\ref{fig4}a illustrates using a 3-generator system that such curved paths exist and can connect $\boldsymbol{\upbeta}_{=}$ to the (unique) global minimum, which we denote by $\boldsymbol{\upbeta}_{\neq}$ as its components are all different.  The corresponding optimal $\lambda^{\max}(\boldsymbol{\upbeta}_{\neq}) \approx -9.41$ represents more than $8$\% improvement over $\lambda^{\max}(\boldsymbol{\upbeta}_{=}) \approx -8.69$.  In general, if a curved path starts at $\boldsymbol{\upbeta}_{=}$, and if $\lambda^{\max}$ decreases monotonically along that path, then it cannot be oriented in an arbitrary direction in the $\boldsymbol{\upbeta}$-space.  We show that it needs to be tangent to a system-specific plane (or hyperplane of co-dimension one for $n > 3$), denoted here by $L$ and defined by the equation $\sum_{i=1}^n u_i v_i \beta_i = 0$, where $u_i$ and $v_i$ are the $i$th component of the left and right eigenvectors, respectively, associated with the eigenvalue $\alpha_2$.  This result, illustrated by the three example paths in Fig.~\ref{fig4}a, follows from the derivation of a formula for $\lambda^{\max}$ and the full analytical characterization of the stability landscape near $\boldsymbol{\upbeta}_{=}$ (both presented in \hyperlink{supp_note_1}{Supplementary Note 1}).

The curved paths of decreasing $\lambda^{\max}$ are part of the complex structure of the stability landscape.  These paths generally lie on a cusp surface, defined by the property that, at any point on the surface, $\lambda^{\max}$ is non-differentiable and locally minimum along any direction transverse to the surface.  The three paths shown in Fig.~\ref{fig4}a all lie on the same cusp surface, which contains both $\boldsymbol{\upbeta}_{=}$ and $\boldsymbol{\upbeta}_{\neq}$.  The intersection between this cusp surface and the plane $M$ (the one perpendicular to $L$) is the green path of monotonically decreasing $\lambda^{\max}$ shown in Fig.~\ref{fig4}.  In fact, there are infinitely many different paths of decreasing $\lambda^{\max}$ on this cusp surface.  Of these paths, the red and blue paths shown in Fig.~\ref{fig4}a share the additional property of being an intersection between pairs of cusp surfaces.  Each of these paths consists of at least two parts that are intersections between different pairs of cusp surfaces, which explains the kinks observed in Fig.~\ref{fig4} as points at which the curve switches from one intersecting surface to another.
In larger systems, we find that their higher-dimensional $\boldsymbol{\upbeta}$-spaces are sectioned by many entangled cusp hypersurfaces associated with spectral degeneracies (as illustrated in Supplementary Fig.~\ref{fig-high-dim-cusps} using the four larger systems we will introduce below).  Their intersections, which themselves form cusp hypersurfaces of lower dimensions, are expected to contain curved paths of monotonically decreasing $\lambda^{\max}$.
The existence of kinks and cusp surfaces in the stability landscape, which makes numerical search for global optima challenging, is not unique to power grids nor phase oscillator networks.  It is a consequence of a much more general mathematical observation that the largest real part of the eigenvalues of a matrix (known as the spectral abscissa), such as $\lambda^{\max}$ we consider here, is a non-smooth, non-convex, and non-Lipschitz function of the matrix elements \cite{Burke:01}.

\begin{figure*}
\centering
\includegraphics[width=1.0\linewidth]{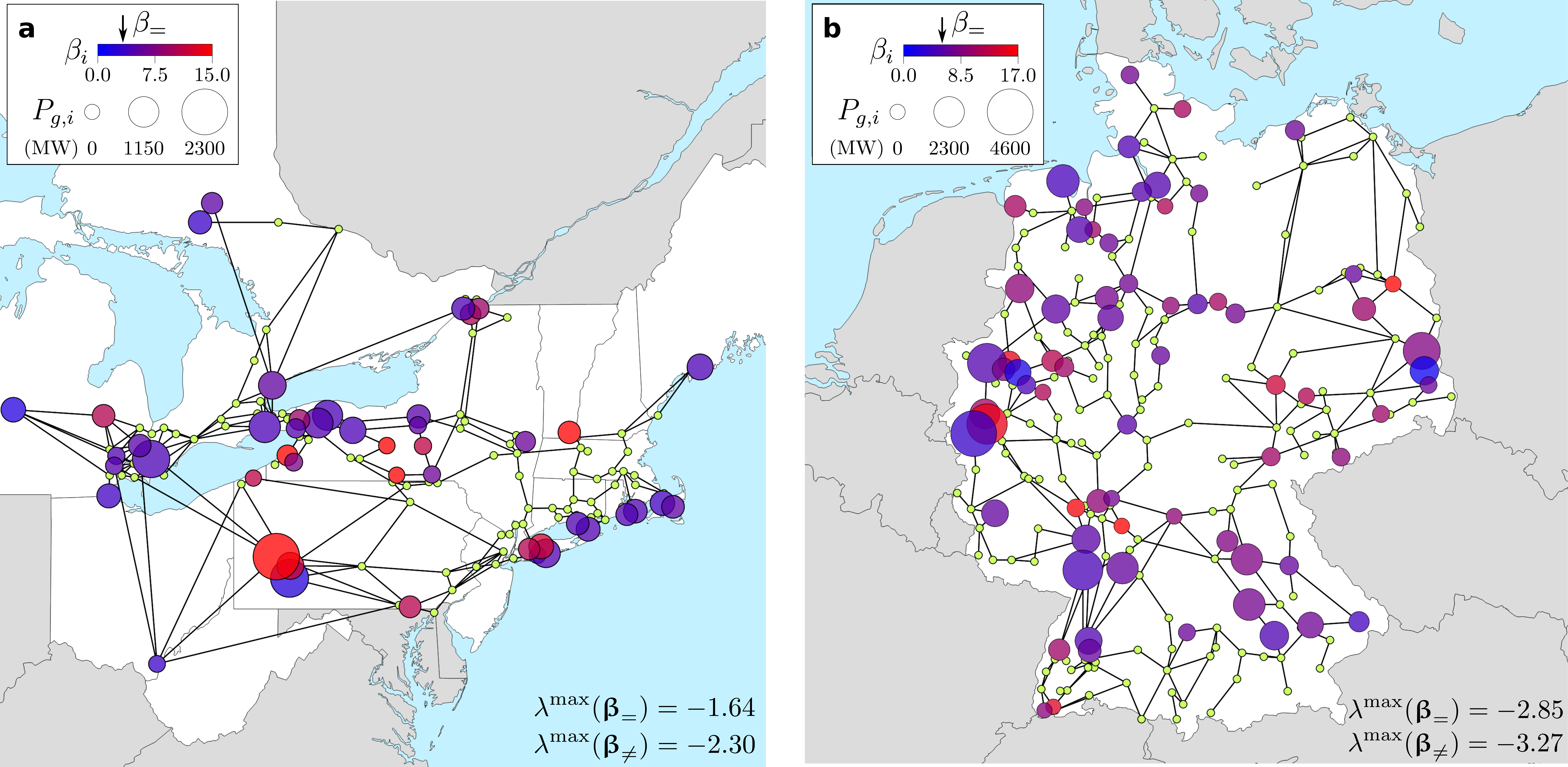}
\caption{\textbf{Heterogeneity of optimized generator parameters $\boldsymbol{\upbeta_i}$ for two power grids.}  \textbf{a}~Portion of the North American power grid corresponding to the former Northeast Power Coordinating Council (NPCC) region.  \textbf{b} German portion of the European power grid.  In both panels, a color-coded circle represents a generator or an aggregate of generators (see Methods for the aggregation procedure used), with the color indicating the corresponding optimal $\beta_i$ in the vector $\boldsymbol{\upbeta}_{\neq}$.  The arrows above the color bars indicate $\beta_{=}$, the optimal uniform value of $\beta_i$.  The $\lambda^{\max}$ values for the uniform and non-uniform optimal $\beta_i$ (in the vectors $\boldsymbol{\upbeta}_{=}$ and $\boldsymbol{\upbeta}_{\neq}$, respectively) are indicated at the bottom of each panel.  The radius of the circle is proportional to the real power output of the generator in megawatts (MW).  Small green dots indicate non-generator nodes.  Details on these systems, including data sources, can be found in Methods.}
\label{fig6}
\end{figure*}

\paragraph{Stabilizing heterogeneity in real power grids.}
\addcontentsline{toc}{subsection}{Stabilizing heterogeneity in real power grids}
Having established that heterogeneous $\boldsymbol{\upbeta}_{\neq}$ can improve stability over the homogeneous $\boldsymbol{\upbeta}_{=}$ for a small example system, we now show that this result extends to much larger, real-world power grids.  Specifically, we study the $48$-generator NPCC portion of the North American power grid and the $69$-generator German portion of the European power grid.
Assessing the stability against small perturbations based on Eqs.~\eqref{eq-en}--\eqref{eqn:P} has the advantage of reducing the complexity of these systems to a single matrix $\mathbf{P}$ and its eigenvalues.
For each system, we identify a local minimum $\boldsymbol{\upbeta}_{\neq}$ that has heterogeneous $\beta_i$ (and thus is distinct from $\boldsymbol{\upbeta}_{=}$) and achieves the lowest $\lambda^{\max}$ over $200$ independent runs of simulated annealing.  We find simulated annealing to be more effective than other methods in locating a minimum on a non-differentiable landscape \cite{Nishikawa:2015}.  The resulting stability improvement over $\boldsymbol{\upbeta}_{=}$ is substantial: $\lambda^{\max}(\boldsymbol{\upbeta}_{=}) - \lambda^{\max}(\boldsymbol{\upbeta}_{\neq}) = 0.66$ for the NPCC network and $\lambda^{\max}(\boldsymbol{\upbeta}_{=}) - \lambda^{\max}(\boldsymbol{\upbeta}_{\neq}) = 0.42$ for the German network.  The optimized $\beta_i$ assignment in $\boldsymbol{\upbeta}_{\neq}$ exhibits substantial heterogeneity across each network and also across the corresponding geographical area (Fig.~\ref{fig6}).

\begin{figure*}
\centering
\includegraphics[width=1.0\linewidth]{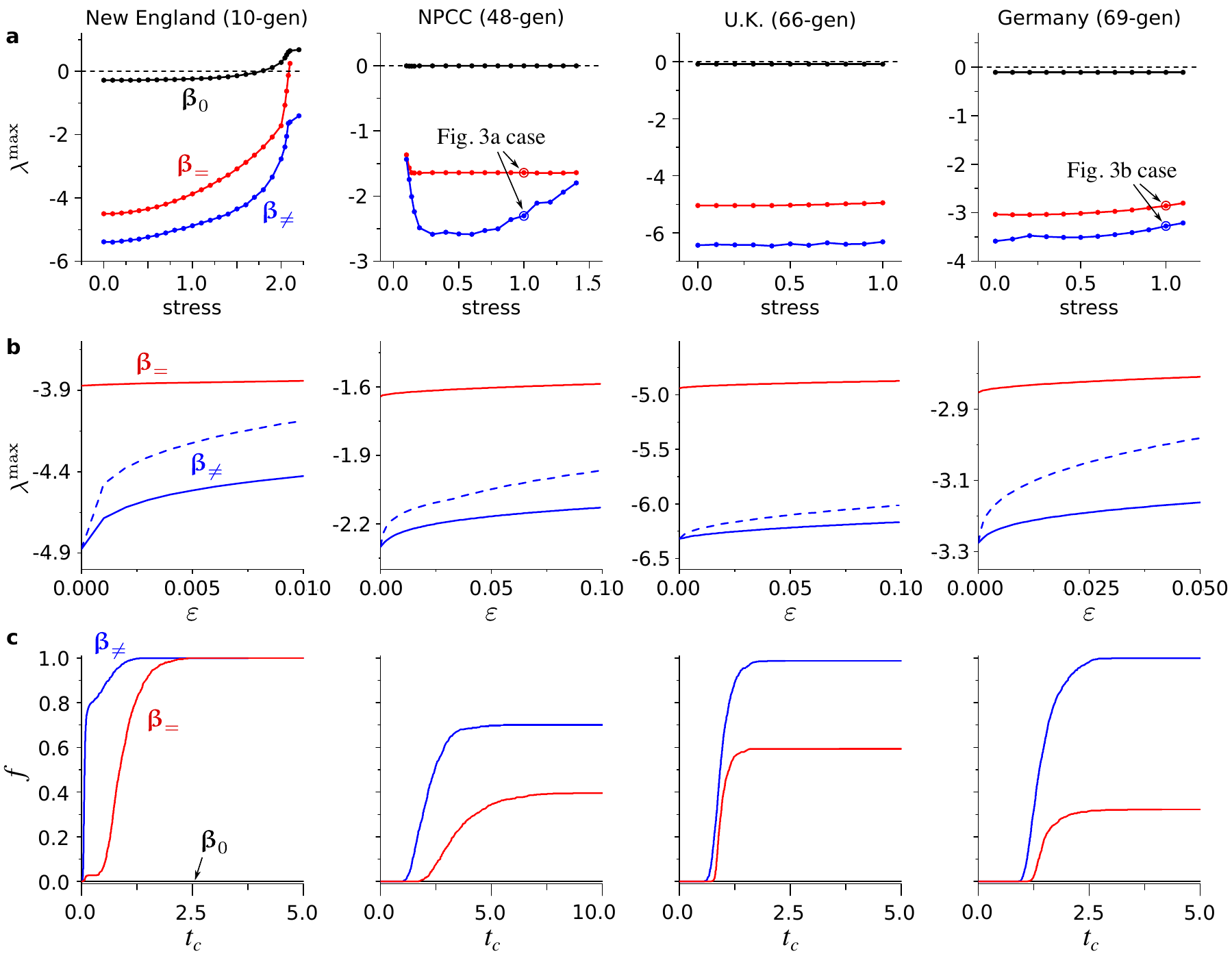}
\caption{\textbf{Improving the stability of power grids with heterogeneity in $\boldsymbol{\upbeta}$.}  The columns correspond to the four systems we consider.  \textbf{a}~Improved Lyapunov exponent $\lambda^{\max}$ as functions of the 
system stress level 
for the heterogeneous optimum $\boldsymbol{\upbeta}_{\neq}$ (blue), the homogeneous optimum $\boldsymbol{\upbeta}_{=}$ (red), and the original parameter $\boldsymbol{\upbeta}_0$ (black).  The cases shown in Fig.~\ref{fig6} are indicated in the second and the last plot.  \textbf{b}~Change of $\lambda^{\max}$ under perturbations of size $\varepsilon$ applied to $\boldsymbol{\upbeta}_{\neq}$ (blue).  We show $\lambda^{\max}$ as a function of $\varepsilon$, where solid and dashed curves indicate the average and the maximum, respectively, over perturbations in 1,000 random directions.  Note that the maximum corresponds to the worst case scenario.  For comparison, we also show the average of $\lambda^{\max}$ when $\boldsymbol{\upbeta}_{=}$ is perturbed (red).  \textbf{c}~Fraction $f$ of trajectories that converge to 
synchronous states
before a given cutoff time $t_c$ for $\boldsymbol{\upbeta}_{\neq}$ (blue), $\boldsymbol{\upbeta}_{=}$ (red), and $\boldsymbol{\upbeta}_0$ (black).  Note that $f$ for $\boldsymbol{\upbeta}_0$ remains zero for all $t_c<10$ seconds in all cases.}
\label{fig_stab}
\end{figure*}

To validate the prevalence of such stability-enhancing heterogeneity, we also analyze $\lambda^{\max}$ as a function of \textit{system stress level} (to be precisely defined below) for four different systems, including the two used in Fig.~\ref{fig6} (see Methods for detailed descriptions of the systems and data sources).  To increase or decrease the level of stress in a given system, we scale the power output of all generators and the power demand at all nodes by a common constant factor.  We then re-compute the power flows across the entire network and the parameters of Eq.~\eqref{eq-en}.  The system stress level is then defined to be the common scaling factor used in this procedure.
Thus, a stress level of $1$ for a given system corresponds to the original demand level in the corresponding dataset.
For each stress level, we estimate $\lambda^{\max}$ at $\boldsymbol{\upbeta} = \boldsymbol{\upbeta}_{\neq}$ from $200$ independent simulated annealing runs.  Over the entire range of stress levels considered, we consistently observe a smaller $\lambda^{\max}$ for $\boldsymbol{\upbeta}_{\neq}$ compared to $\boldsymbol{\upbeta}_{=}$, the optimal homogeneous $\beta_i$ assignment, and to $\boldsymbol \beta_0$, the original $\beta_i$ assignment in the dataset (Fig.~\ref{fig_stab}a).

To test the robustness of the identified optimal $\lambda^{\max}$ against uncertainties in the $\beta_i$ values, we study how $\lambda^{\max}$ changes under perturbations along random directions in the ${\boldsymbol \beta}$-space in the vicinity of $\boldsymbol{\upbeta}_{\neq}$ and (for comparison) in the vicinity of $\boldsymbol{\upbeta}_{=}$.  For the stress level of $1$ and for each random direction, we compute $\lambda^{\max}$ as a function of the perturbation size $\varepsilon$, measured in $2$-norm.  The resulting statistics from $1{,}000$ random directions indicate that, for each system, there is a sizable neighborhood of the optimum $\boldsymbol{\upbeta}_{\neq}$ in which $\lambda^{\max}$ is significantly lower than at $\boldsymbol{\upbeta}_{=}$, representing a stability improvement against small perturbations (Fig.~\ref{fig_stab}b).

To show that the improvement is also observed for stability against large perturbations, we define a generalized notion of attraction basin as a set of initial conditions whose corresponding trajectories satisfy a criterion for convergence to 
synchronous states (a variation of the so-called basin stability~\cite{bst}).  Here, the convergence criterion we use is that the instantaneous frequency enters into a narrow band around $\omega_\mathrm{s}$ (within $\pm 0.3$\,Hz) and remains inside the band until $t_{\max}=10$ seconds.  This criterion is similar to what is typically used for transient stability analysis in power system engineering.  It also captures a variety of 
synchronous states,
including not only those corresponding to fixed points of Eq.~\eqref{eq-en} (with constant phase angle differences), but also those corresponding to time-dependent solutions of Eq.~\eqref{eq-en}.  To account for large perturbations, we consider initial conditions with arbitrary phase angles and frequencies within $1$ Hz of the nominal frequency ($60$ Hz for the New England and NPCC systems; $50$ Hz for the U.K. and German systems).
Each initial condition can be regarded as resulting from a large impulse-like disturbance, such as a disconnection of a significant portion of the grid or a system-wide demand surge.
The size of the basin can then be quantified using the fraction $f$ of the corresponding trajectories that converge before a given cutoff time $t_c$, i.e., the fraction of those that satisfy $\lvert \dot\delta_i(t) \rvert / (2\pi) \le 0.3$~Hz for all $t \in [t_c, t_{\max}]$.  For each $t_c$, the fraction $f$ is estimated using $1{,}000$ initial conditions sampled randomly and uniformly from all states satisfying the criteria described above.  As shown in Fig.~\ref{fig_stab}c, we find that the estimated $f$ is significantly larger for $\boldsymbol{\upbeta}_{\neq}$ than for $\boldsymbol{\upbeta}_{=}$ (which in turn is much larger than for $\boldsymbol{\upbeta}_0$). 
This indicates that the likelihood for the system to return to stable operation after a large disturbance is higher for the heterogeneous optimal $\beta_i$ than for the homogeneous optimal ones.
We also observe that larger systems tend to exhibit larger increase in the size of the asymptotic basins (i.e., in the value of $f$ for $t_c \to \infty$).

\begin{figure*}
\centering
\includegraphics[width=1.0\linewidth]{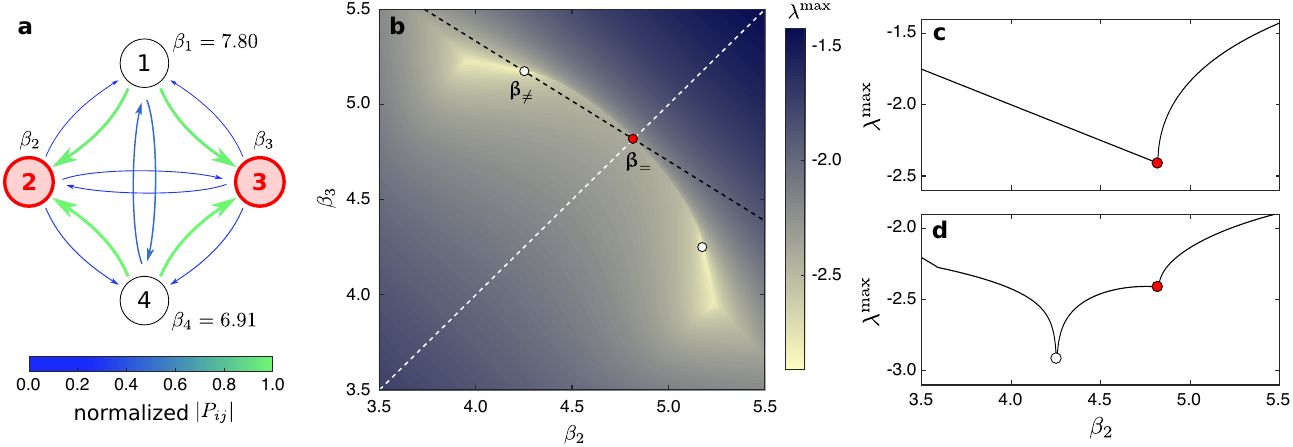}
\vspace{-7mm}
\caption{\textbf{Illustrating CSB in a 4-generator example system.}  \textbf{a}~Effective interaction network connecting the generators and given by matrix $\mathbf{P}$.  Both the thickness and color of the arrow connecting node $j$ to node $i$ represent the interaction strength $|P_{ij}|$, with the thickness proportional to $|P_{ij}|$ and the color encoding $|P_{ij}|$ normalized by its maximum over all $i$ and $j$ with $i \neq j$.  \textbf{b}--\textbf{d}~Dependence of $\lambda^{\max}$ on $\beta_2$ and $\beta_3$, with the values of $\beta_1$ and $\beta_4$ set to the values indicated in \textbf{a}.  In \textbf{b}, $\lambda^{\max}$ is color coded to visualize the full 2D landscape.  In \textbf{c}, $\lambda^{\max}$ is shown as a function of $\beta_2$ along the white dashed line in \textbf{b}, corresponding to $\beta_2 = \beta_3$.  It attains its minimum value $\approx$\,$-2.40$ at $\beta_2 = \beta_3 \approx 4.80$ (red circle).  In \textbf{d}, $\lambda^{\max}$ as a function of $\beta_2$ along the black dashed line in \textbf{b} attains its minimum value $\approx$\,$-2.97$ at $(\beta_2, \beta_3) \approx (4.27, 5.17)$ (white circle).  Thus, the substantially improved optimal $\lambda^{\max}$ is possible only when the permutation symmetry between generators 2 and 3 is broken.  For details on the system, see Methods.}
\label{fig_small_example}
\end{figure*}

\paragraph{Isolating converse symmetry breaking.}
\addcontentsline{toc}{subsection}{Isolating converse symmetry breaking}
Since real power systems generally have heterogeneity in $a_i$, $c_{ik}$, and $\gamma_{ik}$, the stability improvement enabled by the $\beta_i$ heterogeneity (and the associated system asymmetry) could in principle be a compensation for heterogeneity in the network structure, power demand and generation, or other component parameters (and the associated system asymmetries).  To illustrate that no such compensation is needed and that CSB can be responsible for stability improvement, we use an example system consisting of four generators connected to each other and to one load (see Supplementary Fig.~\ref{fig_small_example_syst_diagram} for a system
diagram).  This system is symmetric with respect to the permutation of generators 2 and 3 if $\beta_2 = \beta_3$, and this symmetry is reflected in the property that $P_{2j} = P_{3j}$ for all $j$ in the corresponding interaction matrix $\mathbf{P}$ (Fig.~\ref{fig_small_example}a).
The minimum $\lambda^\mathrm{max}$ possible for this symmetric system is $\lambda^\mathrm{max} \approx -2.40$, which can be decreased further by more than 20\% to $\lambda^\mathrm{max} \approx -2.97$ if the $\beta_2 = \beta_3$ constraint is lifted (Fig.~\ref{fig_small_example}b--d).
This demonstrates CSB for this system under a range of noise levels: breaking the system's symmetry under the permutation of generators 2 and 3 is required for $\lambda^\mathrm{max}$ to cross the stability threshold and make the (symmetric) 
synchronous state
stable.
We note that the observation of CSB depends on the system's symmetry.  While CSB is observed in this 4-generator system (with a two-generator permutation symmetry), we do not observe CSB in a variant of the system with the four-generator permutation symmetry.
We also note that,
while the optimal $\beta_i$ assignment does not 
share the two-generator permutation of the system,
the two-dimensional stability landscape does, and it features a pair of equally optimal assignments related to each other through the symmetry (Fig.~\ref{fig_small_example}b).  It is instructive to compare this result with the mass-spring system in Fig.~\ref{fig_mass_spring_example}, where similar breaking of a permutation symmetry (between masses 1 and 3) for a symmetric landscape (where optimal $b_1$ and $b_3$ are necessarily different but can be swapped) is shown to underlie optimal damping.

\begin{figure*}
\centering
\includegraphics[width=1.0\linewidth]{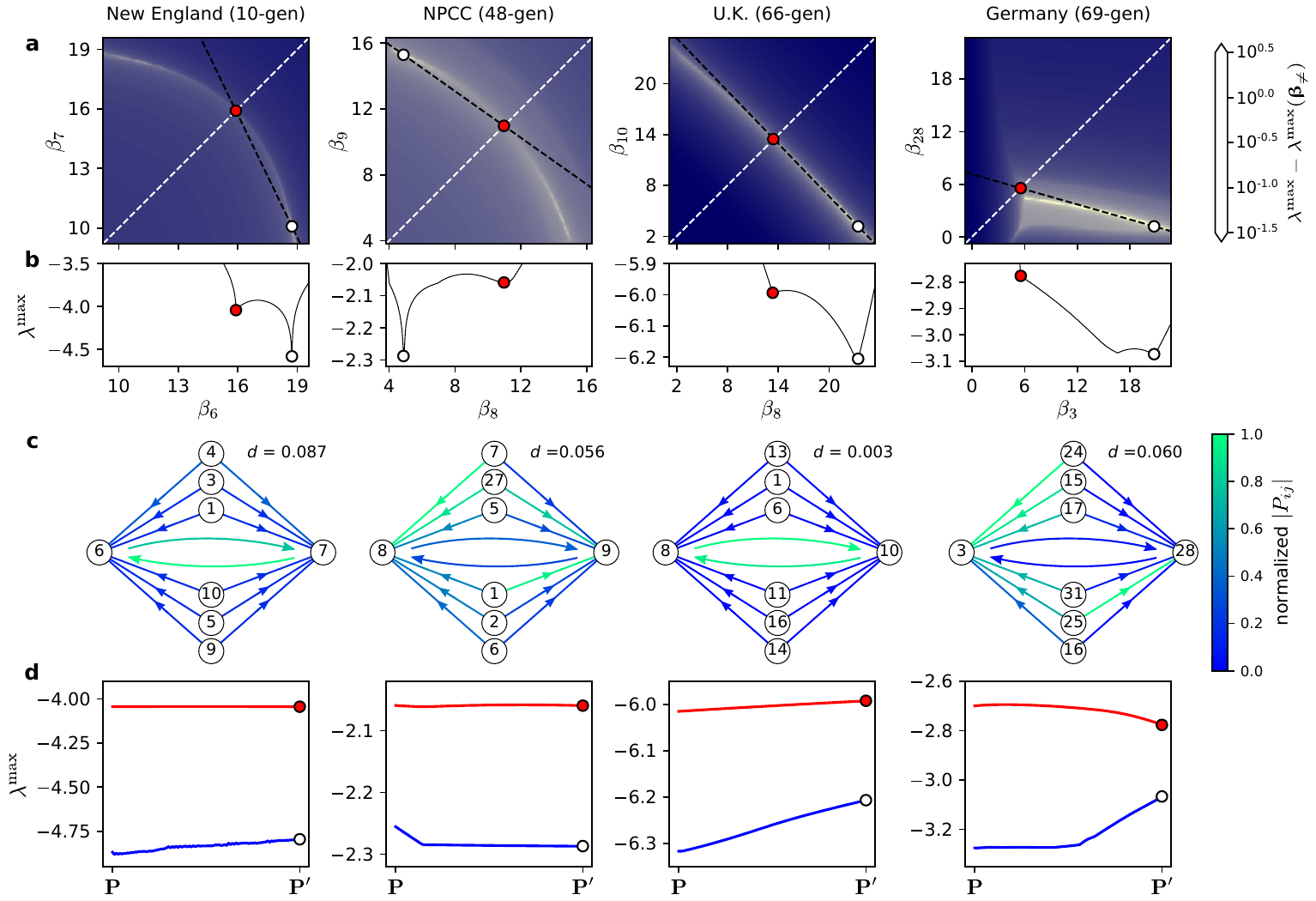}
\caption{\textbf{Isolating the CSB effect in power grids.}  Each column synthesizes results for the system indicated at the top.  \textbf{a} 2D stability landscape $\lambda^{\max}(\beta_{i_1}, \beta_{i_2})$, where $i_1$ and $i_2$ are the nodes whose permutation holds the symmetrized matrix $\mathbf{P}'$ invariant.  In each panel, the red circle marks the optimal $(\beta_{i_1}, \beta_{i_2})$ on the diagonal $\beta_{i_1} = \beta_{i_2}$ (white dashed line), while the white circle marks the optimal when $\beta_{i_1} \neq \beta_{i_2}$ is allowed.  The other $\beta_i$ are fixed at the values in $\boldsymbol{\upbeta}_{\neq}$ identified for a stress level of $1$ in Fig.~\ref{fig_stab}.  \textbf{b} Stability $\lambda^{\max}$ as a function of $\beta_{i_1}$ along the black dashed line connecting the red and white circles in \textbf{a}, respectively.  \textbf{c} Input strength patterns of the nodes $i_1$ and $i_2$ for the original matrix $\mathbf{P}$.  The color of each arrow indicates $\vert P_{ij} \vert$, normalized by the largest value of $\vert P_{ij} \vert$ shown.  For nodes $i_1$ and $i_2$, we show incoming links from the top six common neighbors in terms of the input strength.  Also shown is the distance $d$ from the original matrix $\mathbf{P}$ to its symmetrized version $\mathbf{P}'$, in which the two nodes receive identical incoming (weighted) links, defined as $d \equiv \Vert\mathbf{P} - \mathbf{P}'\Vert_2 / \Vert\mathbf{P}\Vert_2$, where $\Vert\cdot\Vert_2$ denotes the matrix 2-norm.  \textbf{d} Change in the optimal $\lambda^{\max}$ with (red) and without (blue) the constraint $\beta_{i_1} = \beta_{i_2}$, as we interpolate between the original matrix $\mathbf{P}$ and its symmetrized version $\mathbf{P}'$.  Node indexing is described for all four systems in Methods.}
\label{fig_2node}
\end{figure*}

Having established that $\beta_i$ heterogeneity alone can enhance stability through CSB, we now introduce a systematic method to separate CSB from other mechanisms that involve interplays between multiple heterogeneities.  For this purpose, we transform matrix $\mathbf{P}$ for each system used in Fig.~\ref{fig_stab}, which does not have a pairwise node permutation symmetry, to a slightly different matrix $\mathbf{P}'$ that does have the symmetry.  More precisely, for a given pair of nodes $i_1$ and $i_2$, we define this symmetrized matrix $\mathbf{P}'$ by $P'_{i_1 j} = P'_{i_2 j} \equiv (P_{i_1 j} + P_{i_2 j})/2$ for all $j \neq i_1$ nor $i_2$, making it symmetric under the permutation of nodes $i_1$ and $i_2$.  To elucidate CSB for each system, we chose a node pair that simultaneously minimizes the difference between $\mathbf{P}$ and $\mathbf{P}'$ and maximizes the amount of stability improvement observed for $\mathbf{P}'$.  For the four systems in Fig.~\ref{fig_stab}, the stability of the symmetrized system can clearly be enhanced by allowing $\beta_{i_1} \neq \beta_{i_2}$, and much of the enhancement is maintained as one interpolates from the symmetrized system back to the original system (Fig.~\ref{fig_2node}).  This indicates that a significant portion of the stability enhancement for the original system can be attributed to CSB.

\section*{Discussion}
\addcontentsline{toc}{section}{Discussion}

Our demonstration that heterogeneity of generators can enhance the stability of 
synchronous states
in a range of power grids suggests that there is large previously under-explored potential for tuning and upgrading current systems for better stability.
Since larger conventional generators have larger inertia and thus larger impact on the stability of other generators, tuning of their parameters may be particularly beneficial.
While we focused on the heterogeneity of a specific generator parameter here, further stability enhancement is likely to be possible by exploiting heterogeneity in other generator parameters and in the parameters of other network components as well as in the network topology.
We suggest that such stability enhancement opportunities exist beyond power systems and extend to any network whose function benefits from homogeneous dynamics and whose stability depends on tunable system parameters.
For example, the results presented here suggest that CSB can potentially be observed for coupled oscillatory flows in microfluidic networks and for networks of coupled chemical reactors whose oscillatory node dynamics is close to a Hopf bifurcation.
It is known \cite{TakashiAIS} that such systems can be parameterized so that their Jacobian matrices take a form generalizing Eq.~\eqref{eqn:J}, and they are thus capable of exhibiting 
CSB.
The approach we developed here to isolate CSB is versatile and can be applied broadly to systems for which different heterogeneities co-occur.
Determining how prevalent CSB is and how it depends on the properties of the system (e.g., the network size, link distribution, and node dynamics) are important questions for future research.

It is instructive to interpret our results and contrast them with past approaches in network optimization.  In seeking the best approach, one may form two complementary hypotheses.  One hypothesis, invoked in the past, was that the stability of the desired homogeneous states would be optimal when the system is homogeneous; the approach would thus be to limit the optimization search to the low-dimensional parameter subspace corresponding to networks with identical parameter values for all nodes.  The other hypothesis, validated here, is that optimal stability of the desired homogeneous states is generally obtained with heterogeneous parameter assignment, which implies that the search for this optimum requires exploring the high-dimensional parameter space without making {\it a priori} assumptions on how the parameters of different nodes are related.  Realizing this can lead to new control approaches designed to manipulate these parameters for further optimization of stability.  We suggest that the fresh opportunities for network optimization and control revealed in this study apply to network systems in general and thus have the potential to inspire new discoveries in many different disciplines.

\section*{Methods}
\addcontentsline{toc}{section}{Methods}

\paragraph{Power-grid datasets.}
\addcontentsline{toc}{subsection}{Power-grid datasets}
Here, we describe the sources of data for the six power-grid networks considered (the 3-generator system in Fig.~\ref{fig4}; the New England, NPCC, U.K.,
and German systems in Figs.~\ref{fig6}, \ref{fig_stab}, and \ref{fig_2node}; and the 4-generator system in Fig.~\ref{fig_small_example}).  For each system, the data provide the net injected real power at all generator nodes, the power demand at all non-generator nodes, and the parameters of all power lines and transformers.  These parameters are sufficient to determine all active and reactive power flows in the system using a standard power flow calculation.  The data also provide the generators' dynamic parameters $H_i$, $D_i$, and $x_{\mathrm{int},i}$ used in our stability calculations.
The parameters $H_i$ and $D_i$ are the inertia and damping constants, respectively, that define the effective damping parameter through the relation $\beta_i = D_i/(2H_i)$.  The parameter $x_{\mathrm{int},i}$ represents the internal reactance of generator $i$ and is used in the calculation of the parameters $a_i$, $c_{ij}$, and $\gamma_{ij}$.  In each system, nodes are indexed as in the original data source (except for the German power grid; see below).

\begin{itemize}
\item\textbf{3-generator test system (3-gen):} For this IEEE 3-generator, 9-node test system, which appeared in Ref.~\citen{And:03}, we used the data file (data3m9b.m) available in the PST toolbox~\cite{PST}.  This system represents the Western System Coordinating Council (WSCC), which was part of the region now called the Western Electricity Coordinating Council (WECC) in the North American power grid.  The data file provides all necessary dynamical parameters for each generator.
\item\textbf{New England test system (10-gen):} For the IEEE 10-generator, 39-node test system, as described in Refs.~\citen{Pai:89} and \citen{Athay:79}, we used the data file (case39.m) available in the MATPOWER toolbox~\cite{matpower}, with dynamic parameters added manually from Ref.~\citen{Pai:89}.  This is a reduced model representing the New England portion of the Eastern Interconnection in the North American power grid, with one generator representing the connection to the rest of the grid.
\item\textbf{NPCC power grid (48-gen):} For the 48-generator, 140-node NPCC power grid~\cite{Chow}, we used the data file (data48em.m) available in the PST toolbox~\cite{PST}.  The system represents the former NPCC region of the Eastern Interconnection in the North American power grid and includes an equivalent generator/load node representing the rest of the Interconnection.  The data file provides $H_i$ and $x_{\mathrm{int},i}$ for all generators (while it assumes $D_i=0$).  We generated $D_i$ randomly by sampling from the uniform distribution on the interval $[1,3]$ (in per unit on the system base, as specified by the data file).  The geographic coordinates of the nodes used in Fig.~\ref{fig6}a were extracted from Ref.~\citen{Junjian:16}, and the coastline and boundary data used to draw the map were obtained from Natural Earth~\cite{web-earth}.
\item\textbf{U.K. power grid (66-gen):} For the 66-generator, 29-node U.K. power grid, we used the data file (GBreducednetwork.m) available from Ref.~\citen{web-UK}.  The system represents a reduced model for the power grid of Great Britain.  The dynamical parameters, $H_i$, $D_i$, and $x_{\mathrm{int},i}$, were generated randomly by sampling from the uniform distribution on the intervals, $[1,5]$, $[1,3]$, and $[0.001, 0.101]$, respectively.  The generated parameters values for each generator are in per unit on its own machine base, i.e., normalized by the reference values computed from the power base for the generator (chosen to be $1.5$ times the maximum real power generation provided in the data file).  For stability calculation, we converted these values to the corresponding values in per unit on a common system base.
\item\textbf{German power grid (69-gen):} For the 69-generator, 228-node German power grid, we created the data from the ENTSO-E 2009 Winter model~\cite{Bialek:13}.  The ENTSO-E model is a DC power flow model of the continental Europe and contains 1,486 nodes and 565 generators.  We first created a dynamical model for the entire ENTSO-E network by solving the DC power flow and converting it to an AC power flow solution (assuming a $0.95$ power factor at each node), and then generating dynamical parameters using the same method as for the U.K. grid.  For any node with multiple generators attached, the net reactive power injection was distributed among these generators in proportion to their real power generation.  From this full ENTSO-E model, we extracted the German portion by eliminating (using Kron reduction) all the nodes outside Germany (identified using the country label ``D'' representing Germany in the dataset).  We re-indexed the extracted nodes consecutively, preserving the original ordering.  The geographic coordinates of the nodes used in Fig.~\ref{fig6}b were extracted from the PowerWorld data files available from Ref.~\citen{Bialek:13}, and the coastline and boundary data used to draw the map were obtained from Natural Earth~\cite{web-earth}.
\item\textbf{4-generator example system:} For the 4-generator, 5-node example system used in Fig.~\ref{fig_small_example}, we show a full system diagram in Supplementary Fig.~\ref{fig_small_example_syst_diagram}, indicating the main parameters of the components.  When the damping parameters of generators $2$ and $3$ are equal (i.e., $\beta_2 = \beta_3$), the system is symmetric under the permutation of these generators.  MATLAB code for running simulations on this system, which includes the full set of parameters and uses the MATPOWER toolbox~\cite{matpower}, is available from our GitHub repository~\cite{github}.
\end{itemize}

\paragraph{Aggregation of generators and effective damping parameter $\boldsymbol{\upbeta_i}$.}
\addcontentsline{toc}{subsection}{Aggregation of generators and effective damping parameter beta i}

If a subset of generators are synchronized in the sense that $\delta_i - \delta_j$ is constant in time for any two generators $i$ and $j$ in the subset, then they can be represented by a single equivalent generator using a Zhukov-based aggregation method similar to that described in Ref.~\citen{Chow}.  In this method, the equivalent generator has inertia constant $\sum_i H_i$ and damping constant $\sum_i D_i$, where the sums are taken over the generators $i$ in the subset.  The effective damping parameter of the equivalent generator is then $\sum_i D_i/(2\sum_i H_i) = \bar{D}/(2\bar{H})$, where $\bar{D}$ and $\bar{H}$ are respectively the average of the inertia and damping constants of the generators in the subset.  Thus, the aggregation does not introduce any artifactual heterogeneity.

\section*{Data Availability}
Data on all six systems we consider (described in Methods) and detailed data of the core results presented in the figures are available from our GitHub repository \cite{github}.

\section*{Code Availability}
Essential code for reproducing the core results in all figures, as well as scripts for generating plain versions of the figures, is available from the GitHub repository \cite{github}.

\vspace{3mm}

\section*{Acknowledgements}
The authors thank Alex Mercanti and Yuanzhao Zhang for insightful discussions.
This research was supported by Northwestern University's Finite Earth Initiative (supported by Leslie and Mac McQuown) and ARPA-E Award No.~DE-AR0000702 and benefited from logistical support provided by Northwestern University's Institute for Sustainability and Energy.

\section*{Author contributions}
F.M., T.N., and A.E.M.\ designed the research, analyzed the results, and wrote the paper.  F.M.\ performed the simulations.  All authors approved the final manuscript.

\section*{Competing interests}
The authors declare no competing interests.

\section*{Additional information}
\textbf{Correspondence} and requests for material should be addressed to T.N.

\clearpage
\baselineskip18pt
\setcounter{page}{1}
\setcounter{equation}{0}
\renewcommand{\theequation}{S\arabic{equation}}

\noindent\textbf{\LARGE Supplementary Information}\\[1mm]
\noindent\textit{\mytitle}
\addcontentsline{toc}{section}{Supplementary Information}

\hoffset 0cm
\textwidth 16cm
\bigskip\bigskip
\noindent\textbf{\Large Supplementary Figures}
\newcounter{sfigure}
\renewcommand{\figurename}{Supplementary Fig.}
\renewcommand{\thefigure}{\arabic{sfigure}}
\addcontentsline{toc}{subsection}{Supplementary Figures}

\vspace{15mm}

\addtocounter{sfigure}{1} 
\begin{figure}[ht]
\centering
\includegraphics[width=3in]{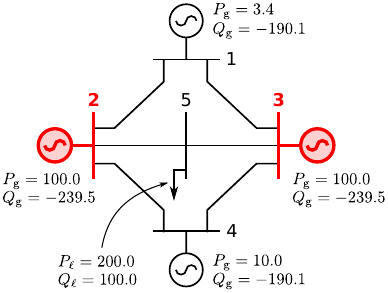}
\vspace{5mm}
\caption{
\textbf{System diagram for the 4-generator example system in Fig.~\ref{fig_small_example}.}  
The generators at nodes 1--4 produce active and reactive power $P_g$ and $Q_g$ (in MW and MVAR, respectively).  The load at node 5 consumes active and reactive power $P_{\ell}$ and $Q_{\ell}$.  The other generator parameters are identical among nodes 1--4, except for the tunable damping parameters $\beta_i$, while the power line parameters are identical among all six lines connecting the nodes.}
\label{fig_small_example_syst_diagram}
\end{figure}

\addtocounter{sfigure}{1} 
\begin{figure}[ht!]
\centering
\includegraphics[width=1.0\textwidth]{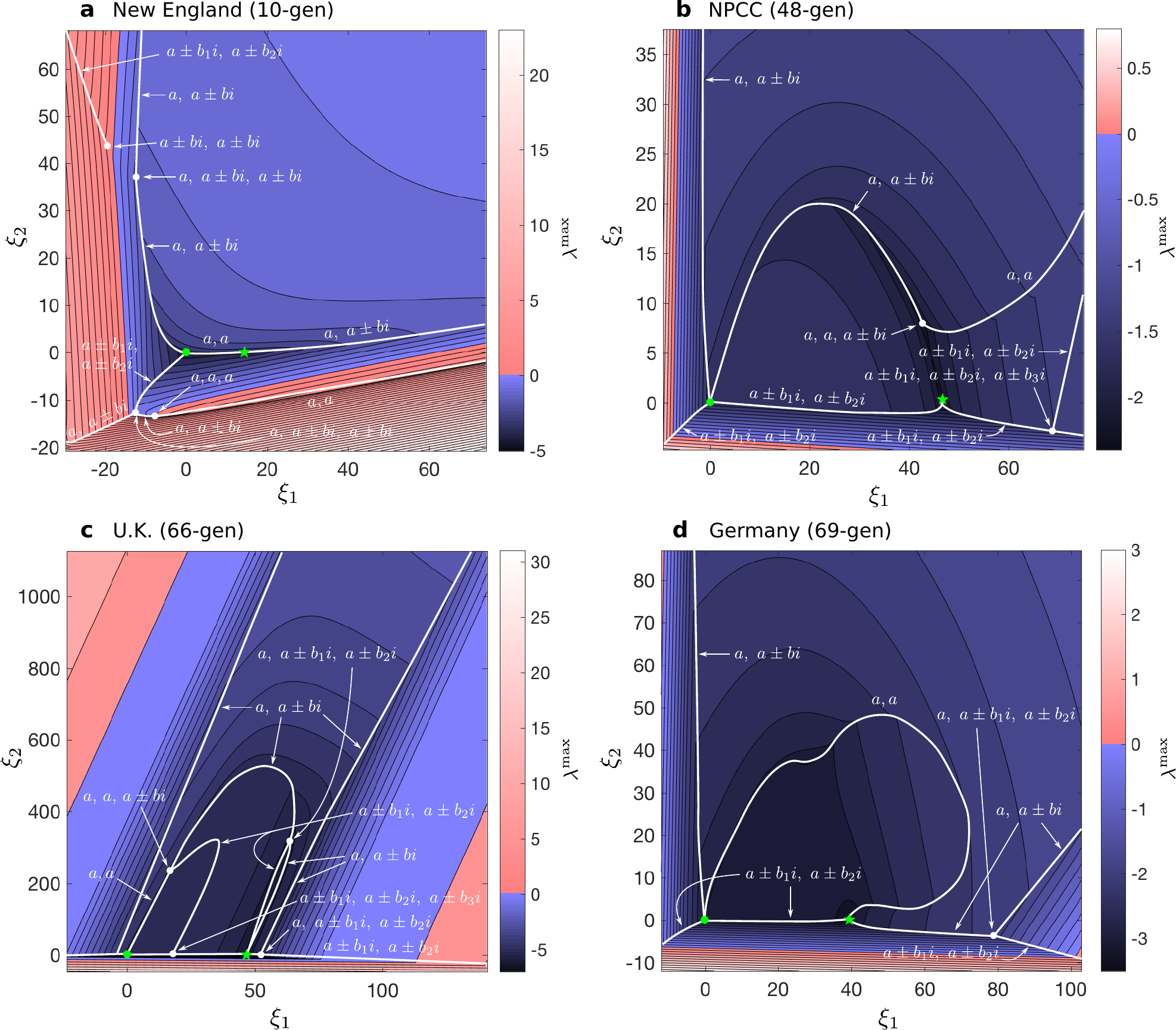}
\vspace{-5mm}
\caption{\textbf{Cusp hypersurfaces in the stability landscape of larger networks.}  
\textbf{a}--\textbf{d}~Contour levels of $\lambda^{\max}$ on the plane $M$, defined as the hyperplane perpendicular to $L$ that contains $\boldsymbol{\upbeta}_{=}$ and $\boldsymbol{\upbeta}_{\neq}$ (as in Fig.~\ref{fig4}), for the four systems used in Fig.~\ref{fig_2node}.  The horizontal coordinate $\xi_1$ represents the Euclidean distance from the point $\boldsymbol{\upbeta}_{=}$ along the line connecting $\boldsymbol{\upbeta}_{=}$ to $\boldsymbol{\upbeta}_{\neq}$, and $\xi_2$ is the distance along the line orthogonal to the $\xi_1$-axis.
A white curve indicates the (one-dimensional) cross section of a codimension-one cusp hypersurface and corresponds to single degeneracy of the real parts of the eigenvalues of the Jacobian $\mathbf{J}$, such as two identical real eigenvalues $\{a,a\}$, two pairs of complex conjugate eigenvalues with matching real parts $\{a \pm b_1 i,\ a \pm b_2 i\}$, and one real eigenvalue matching the real parts of a conjugate eigenvalue pair $\{a,\ a \pm b i\}$.
The green dot and star indicate $\boldsymbol{\upbeta}_{=}$ and $\boldsymbol{\upbeta}_{\neq}$, respectively.
These two points and the white dots correspond to the cross sections of cusp hypersurfaces of higher codimensions, each representing specific double or higher degeneracy of the real parts of the eigenvalues.}
\label{fig-high-dim-cusps}
\end{figure}

\clearpage

\baselineskip15pt
\noindent{\bf\Large Supplementary Notes}
\renewcommand{\thesubsubsection}{\arabic{subsubsection}}

\vspace{11mm}
\hypertarget{supp_note_1}
\noindent{\bf\large Supplementary Note 1: Analysis of the stability landscape around $\boldsymbol{\upbeta_{=}}$}
\addtocounter{section}{1}
\addcontentsline{toc}{subsection}{Supplementary Note 1}

\vspace{6mm}\noindent
Our detailed analysis of the stability landscape around the point $\boldsymbol{\upbeta}_{=}  = (\beta_{=},\ldots,\beta_{=})$, $\beta_{=} = 2\sqrt{\alpha_2}$, is divided into three parts.  We first derive a formula for the maximum Lyapunov exponent $\lambda^{\max}$ in terms of the coefficients of the characteristic polynomial in a neighborhood of $\boldsymbol{\upbeta_{=}}$, which allows us to express a necessary condition for $\lambda^{\max}$ to decrease along a given path (Sec.~\ref{si-subsec-A}).  We then derive equations that relate these coefficients to $\boldsymbol{\upbeta}$ (Sec.~\ref{si-subsec-B}).  Finally, we apply the Implicit Function Theorem (IFT) to these equations and show that, for a generic system, any descending path must be tangent to a system-specific hyperplane in the $\boldsymbol{\upbeta}$-space (Sec.~\ref{si-subsec-C}).  In that section, we also derive an explicit formula for $\lambda^{\max}$ in terms of the parameterization of the path when the path is transverse to the hyperplane.  Two example cases, $n=2$ and $n=3$, are presented for illustration in Sec.~\ref{si-subsec-D}.

\subsubsection[1. Maximum Lyapunov exponent on arbitrary path through homogeneous beta]{Maximum Lyapunov exponent $\boldsymbol{\lambda^{\max}}$ on arbitrary path through $\boldsymbol{\upbeta_{=}}$}
\label{si-subsec-A}

In the main text, we introduced the $2n \times 2n$ Jacobian matrix $\mathbf{J}$ that characterizes the linearized dynamics of the generators. Recall that 
\begin{equation}
\mathbf{J}=\begin{pmatrix}
\,\,\, \mathbf{O} & \,\,\, \mathbf{I}\,\,\, \\
-\mathbf{P} & -\mathbf{B}\,\,\,
\end{pmatrix},
\end{equation}
where $\mathbf{O}$ and $\mathbf{I}$ denote the null and the identity matrix of size $n$, respectively, the $n \times n$ matrix $\mathbf{P}=(P_{ik})$ is given by Eq.~\eqref{eqn:P} in the main text, and $\mathbf{B}$ is the diagonal matrix of elements 
$\beta_i$.  We define
\begin{equation}\label{eqn:J-bar}
\overline{\mathbf{J}}=\begin{pmatrix}
\,\,\, \mathbf{O} & \,\,\, \mathbf{I}\,\,\, \\
-\mathbf{P}/\beta_{=}^2 & -\mathbf{B}/\beta_{=}\,\,\,
\end{pmatrix},
\end{equation}
assuming $\beta_{=}>0$ (or, equivalently, $\alpha_2 > 0$).  Note that $\nu$ is an eigenvalue of $\overline{\mathbf{J}}$ if and only if $\beta_{=}\nu$ is an eigenvalue of $\mathbf{J}$, since the characteristic polynomial of $\overline{\mathbf{J}}$ can be written as
\begin{equation}
\begin{split}
\det(\overline{\mathbf{J}} - \nu\mathbf{I}) 
&= \det(\nu^2\mathbf{I} + \nu\mathbf{B}/\beta_{=} + \mathbf{P}/\beta_{=}^2)\\
&= \beta_{=}^{-2n} \det(\beta_{=}^2\nu^2\mathbf{I} + \beta_{=}\nu\mathbf{B} + \mathbf{P})\\
&= \beta_{=}^{-2n} \det(\mathbf{J} - \beta_{=}\nu\mathbf{I}).
\end{split}
\end{equation}

Consider a (possibly curved) path $\boldsymbol{\gamma}$ passing through the point $\boldsymbol{\upbeta}_{=}$ in the space of all ${\boldsymbol \beta}$, parametrized by $\eps$ through a differentiable vector function $\boldsymbol{\upbeta} = \boldsymbol{\gamma}(\eps)$ satisfying $\boldsymbol{\gamma}(0) = \boldsymbol{\upbeta}_{=}$.  We denote the parametrized eigenvalues of $\overline{\mathbf{J}}$ as $\nu_{j_\pm} = \nu_{j_\pm}(\eps)$, $j=1,\ldots,n$.  The eigenvalues of $\textbf{P}$ are denoted by $\alpha_1, \ldots, \alpha_n$, among which we have $\alpha_1 = 0$, and we assume that they are all real, distinct, and indexed so that $0 < \alpha_2 < \cdots < \alpha_n$.  The assumption that these eigenvalues are distinct will be crucial in the arguments that follow.  When $\eps=0$, we have $\mathbf{B} = \beta_{=}\mathbf{I}$, so
\begin{equation}
\det(\overline{\mathbf{J}} - \nu\mathbf{I}) 
= \beta_{=}^{-2n} \det[\beta_{=}^2(\nu^2 + \nu)\mathbf{I} + \mathbf{P}],
\end{equation}
and this implies that $\beta_{=}^2(\nu^2 + \nu) = \alpha_j$, or equivalently $\nu = \bigl(-1 \pm \sqrt{1 - \alpha_j/\alpha_2}\bigr)/2$, whenever $\nu$ is an eigenvalue of $\overline{\mathbf{J}}$.  We thus index $\nu_{j_\pm}(\eps)$ so that 
\begin{equation}\label{eqn:nu_j}
\nu_{j_\pm}(\eps) \xrightarrow{\eps\to 0} \nu_{j_\pm}(0)= \frac{-1 \pm \sqrt{1 - \alpha_j/\alpha_2}}{2}, \quad j = 1,\ldots,n.
\end{equation}
Note that $\nu_{1_+}(\eps) \to \nu_{1_+}(0) = 0$ and $\nu_{1_-}(\eps) \to \nu_{1_-}(0) = -1$ are not relevant for determining the stability of synchronous states since these eigenvalues are associated with perturbation modes that do not affect synchronization.  Thus, the Lyapunov exponent $\lambda^{\max}$ determining the stability is the largest real component among the remaining eigenvalues:  
\begin{equation}\label{eqn:lambda_max}
\lambda^{\max}(\eps) = \beta_{=} \cdot \max_{2 \le j \le n} 
\max \bigl\{ Re(\nu_{j_+}(\eps)), Re(\nu_{j_-}(\eps)) \bigr\}.
\end{equation}
For $\eps = 0$, which corresponds to the point $\boldsymbol{\upbeta}_{=}$, it follows from the formula for $\nu_{j_\pm}(0)$ in Eq.~\eqref{eqn:nu_j} and $\alpha_2 < \alpha_j$, $\forall j \ge 3$, that
\begin{equation}
\lambda^{\max}(0) = \beta_{=} \cdot Re(\nu_{2_+}(0)) = -\frac{\beta_{=}}{2} =  -\sqrt{\alpha_2} = \lambda^{\max}_{=}.
\end{equation}
Since eigenvalues are continuous functions of the matrix elements and $\boldsymbol{\gamma}$ is a continuous function, $\nu_{j_\pm}(\eps)$ changes with $\eps$ continuously.  Thus, for $\eps \neq 0$, the eigenvalue $\nu_{2_+}(\eps)$ determines the maximum in Eq.~\eqref{eqn:lambda_max} for sufficiently small $\eps$, and hence we have
\begin{equation}\label{eqn:lambda_max_eps}
\lambda^{\max}(\eps) = \beta_{=} \cdot Re(\nu_{2_+}(\eps)) = 
\beta_{=} \cdot Re \left(-\frac{c_2}{2} + \frac{1}{2}\sqrt{c_2^2 - 4 d_2} \right),\end{equation}
where we factored the characteristic polynomial of $\overline{\mathbf{J}}$ into quadratic factors as
\begin{equation}\label{char_poly}
\det(\overline{\mathbf{J}} - \nu\mathbf{I}) = \bigl(\nu^2 + c_1\nu\bigr)\bigl(\nu^2 + c_2\nu + d_2\bigr) \cdots \bigl(\nu^2 + c_n\nu + d_n\bigr),
\end{equation}
(noting that $d_1=0$ always holds because $\nu = 0$ is an eigenvalue of $\overline{J}$ associated with $\alpha_1 = 0$).
Equations \eqref{eqn:lambda_max_eps} and \eqref{char_poly} imply that $\lambda^{\max}$ can be viewed as a function of $c_2$ and $d_2$ and thus defines a landscape over the $(c_2,d_2)$-plane, while the path $\boldsymbol{\gamma}(\eps)$ in the $\boldsymbol{\upbeta}$-space corresponds a path on the $(c_2,d_2)$-plane defined by the functions $c_2 = c_2(\eps)$ and $d_2 = d_2(\eps)$.  Defining $f(c_2,d_2) \equiv Re \bigl(-c_2 + \sqrt{c_2^2 - 4d_2}\bigr)/2$, we see that the condition for $\lambda^{\max}$ to be decreasing along the path $\boldsymbol{\gamma}(\eps)$ starting at $\boldsymbol{\upbeta}_{=}$ in the $\boldsymbol{\upbeta}$-space is equivalent to the condition that the corresponding path $(c_2(\eps),d_2(\eps))$ starting at $(c_2,d_2)=(1, 1/4)$ immediately enters the following triangular region of the $(c_2,d_2)$-plane:
\begin{equation}\label{eqn:good_region}
\bigl\{ (c_2,d_2) :\,\, f(c_2,d_2) < -1/2 \bigr\}
= \bigl\{ (c_2,d_2) :\,\, 2(d_2-1/4) > c_2-1 > 0\bigr\}.
\end{equation}
The latter condition implies that the derivatives of $c_2(\eps)$ and $d_2(\eps)$ at $\eps=0$ satisfy $2d'_2(0) \ge c'_2(0)$.  We may assume $c'_2(0) \ge 0$ without loss of generality.  (If not, we simply need to make the change of variable, $\eps \to -\eps$.)  In Sec.~\ref{si-subsec-C}, we show that, for a generic choice of the matrix $\mathbf{P}$, we have $d'_2(0) = 0$.  This implies that, if $\lambda^{\max}$ decreases along the path, we have $c'_2(0) = 0$.  In general, if $c'_2(0) = d'_2(0) = 0$, the nonlinearity of $c_2(\eps)$ and $d_2(\eps)$ determines whether $\lambda^{\max}$ decreases along the path in either direction.  We further show that the condition $c'_2(0) = 0$ translates to the path $\boldsymbol{\gamma}$ being tangent to a specific hyperplane, which we denote by $L$, at the point $\boldsymbol{\upbeta_{=}}$.

\subsubsection[2. Equations relating coefficients to parameter epsilon]{Equations relating coefficients $\boldsymbol{c_i}$ and $\boldsymbol{d_i}$ to parameter $\boldsymbol{\eps}$}
\label{si-subsec-B}

We now derive equations that define the coefficients $c_i$ and $d_i$ implicitly as functions $c_i = c_i(\eps)$ and $d_i = d_i(\eps)$.  We first rewrite the characteristic polynomial of $\overline{\mathbf{J}}$ as
\begin{equation}\label{eqn:ch_poly}
\det(\overline{\mathbf{J}} - \nu\mathbf{I}) 
= \det(\nu^2\mathbf{I} + \nu\mathbf{B}/\beta_{=} + \mathbf{P}/\beta_{=}^2)
= \det(\nu^2\mathbf{I} + \nu\overline{\mathbf{B}} + \mathbf{D}),
\end{equation}
where we used a similarity transformation $\mathbf{Q}^{-1}(\mathbf{P}/\beta_{=}^2)\mathbf{Q} = \mathbf{D}$ and $\mathbf{B}/\beta_{=}$ as $\mathbf{Q}^{-1}(\mathbf{B}/\beta_{=})\mathbf{Q} = \overline{\mathbf{B}}$ (based on the diagonalization of the matrix $\mathbf{P}$), and $\mathbf{D}$ is the diagonal matrix with diagonal elements $\overline{\alpha}_i \equiv \alpha_i/\beta_{=}^2 = \frac{1}{4}\alpha_i/\alpha_2$.  Along the path $\boldsymbol{\gamma}(\eps) = (\gamma_1(\eps), \ldots, \gamma_n(\eps))$, the components of the matrix $\overline{\mathbf{B}}$ can be expressed as 
\begin{equation}\label{eqn:B-tilde}
\overline{B}_{ij}(\eps) = \sum_{\ell=1}^n u_{i\ell} v_{j\ell} \gamma_\ell(\eps)/\beta_{=}, 
\end{equation}
where $u_{i\ell}$ and $v_{i\ell}$ are the $\ell$th component of the left and right eigenvectors of $\mathbf{P}$ associated with the eigenvalue $\alpha_i$, respectively.  We now use the definition of the determinant, 
\begin{equation}
\det(\mathbf{A}) \equiv \sum_{\sigma}\text{sign}(\sigma) \prod_{i=1}^n A_{i\sigma(i)}
= \sum_{\sigma}\text{sign}(\sigma) \cdot A_{1\sigma(1)}\cdots A_{n\sigma(n)},
\end{equation}
where the summation $\sum_{\sigma}$ is taken over all possible permutations $\sigma$ of indices, and $\text{sign}(\sigma)$ is the sign of permutation $\sigma$ (i.e., $\text{sign}(\sigma) = 1$ when $\sigma$ is an even permutation and $\text{sign}(\sigma) = -1$ when $\sigma$ is an odd permutation).
With this definition, we can write the coefficient of the $\nu^k$ term (for $k = 1,\ldots, 2n-1$; no constant term corresponding to $k=0$, since $\nu=0$ is an eigenvalue) of the polynomial in Eq.~\eqref{eqn:ch_poly} as
\begin{equation}\label{eqn:k-coeff}
\sum_{\sigma}\text{sign}(\sigma) \sum_{\{k_i\}} \prod_{i=1}^n E_{i\sigma(i)}^{(k_i)} \cdot \chi \bigl( \textstyle\sum_i k_i = k \bigr),
\end{equation}
where the summation $\sum_{\{k_i\}}$ is taken over all possible combinations of $k_i = 0,1,2$, $i=1,\ldots,n$.  The function $\chi$ is an indicator function defined by $\chi \bigl( \sum_i k_i = k \bigr) = 1$ if $\sum_i k_i = k$ and $\chi \bigl( \sum_i k_i = k \bigr) = 0$ otherwise.  The matrices $E_{ij}^{(k)}$ are defined for $k=0,1,2$ as
\begin{gather}\label{eqn:Eij}
E_{ij}^{(0)} = D_{ij} = \overline{\alpha}_i \delta_{ij},\quad E_{ij}^{(1)} = \overline{B}_{ij}(\eps),\quad E_{ij}^{(2)} = \delta_{ij},
\end{gather}
corresponding to the matrices $\mathbf{D}$, $\overline{\mathbf{B}}$, and $\mathbf{I}$ in Eq.~\eqref{eqn:ch_poly}, respectively ($\delta_{ij}$ is the Kronecker delta function).  Equations~\eqref{eqn:k-coeff} and \eqref{eqn:Eij} provide an expression for the coefficients of $\det(\overline{\mathbf{J}} - \nu\mathbf{I})$ in terms of $\boldsymbol{\upbeta}$, and thus of $\eps$, for a given system and its matrix $\mathbf{P}$.

We now derive a different expression for $\det(\overline{\mathbf{J}} - \nu\mathbf{I})$, this time in terms of the characteristic polynomial coefficients $c_i$ and $d_i$, using Eq.~\eqref{char_poly}.  Ignoring the dependence of $c_i$ and $d_i$ on $\eps$ for the moment and regarding them as independent variables, the coefficient of the $\nu^k$ term can be written as
\begin{equation}\label{eqn:k-coeff2}
\sum_{\{k_i\}} \prod_{i=1}^n a_i^{(k_i)} \cdot \chi \bigl( \textstyle\sum_i k_i = k \bigr),
\end{equation}
where the summation is the same as in Eq.~\eqref{eqn:k-coeff} and 
\begin{gather}\label{eqn:aik-def}
a_i^{(0)} = d_i,\quad a_i^{(1)} = c_i,\quad a_i^{(2)} = 1.
\end{gather}
Setting Eqs.~\eqref{eqn:k-coeff} and \eqref{eqn:k-coeff2} equal to each other, we obtain a set of nonlinear equations that must be satisfied by the variables $c_1,\ldots,c_n$, $d_1,\ldots,d_n$, and $\eps$:
\begin{equation}\label{eqn:relation}
F_k(c_1,\ldots,c_n,d_1,\ldots,d_n,\eps) = 0, \quad k=1,2,\ldots,2n-1,
\end{equation}
or, in vector form, 
\begin{equation}
\mathbf{F}(c_1,\ldots,c_n,d_1,\ldots,d_n,\eps) = \mathbf{0},
\end{equation}
where we have defined $\mathbf{F} \equiv (F_1,\ldots,F_{2n-1})^T$, and the functions $F_k$ are given by
\begin{equation}\label{eqn:Fk}
\begin{split}
F_k &= \sum_{\{k_i\}} \prod_{i=1}^n a_i^{(k_i)} \cdot \chi \bigl( {\textstyle\sum_i k_i = k} \bigr) - \sum_{\sigma}\text{sign}(\sigma) \sum_{\{k_i\}} \prod_{i=1}^n E_{i\sigma(i)}^{(k_i)} \cdot \chi \bigl( {\textstyle\sum_i k_i = k} \bigr) \\
&= \sum_{\{k_i\}} \chi \bigl( {\textstyle\sum_i k_i = k} \bigr) \cdot \Biggl[\prod_{i=1}^n a_i^{(k_i)}  - \sum_{\sigma}\text{sign}(\sigma) \prod_{i=1}^n E_{i\sigma(i)}^{(k_i)} \Biggr].
\end{split}
\end{equation}
This is the equation that implicitly defines the functions $c_i = c_i(\eps)$ and $d_i=d_i(\eps)$.  Note that, when $c_i = 1$, $d_i = \overline{\alpha}_i$, and $\eps=0$ (corresponding to the point $\boldsymbol{\upbeta_{=}}$),  we have $E_{ij}^{(1)} = \overline{B}_{ij}(0) = \delta_{ij}$, and hence
\begin{gather}\label{eqn:beta-tilde}
a_i^{(0)} = d_i(0) = \overline{\alpha}_i,\quad
a_i^{(1)} = c_i(0) = 1,\quad
a_i^{(2)} = 1, \quad E_{ij}^{(k)} = a_i^{(k)} \delta_{ij},
\end{gather}
implying
\begin{align}
F_k(1,\ldots,1,\overline{\alpha}_1,\ldots,\overline{\alpha}_n,0)
&= \sum_{\{k_i\}} \chi \bigl( {\textstyle\sum_i k_i = k} \bigr) \cdot \Biggl[\prod_{i=1}^n a_i^{(k_i)}  - \sum_{\sigma}\text{sign}(\sigma) \prod_{i=1}^n a_{i\sigma(i)}^{(k_i)} \delta_{i\sigma(i)} \Biggr] \nonumber\\
&= \sum_{\{k_i\}} \chi \bigl( {\textstyle\sum_i k_i = k} \bigr) \cdot \Biggl[\prod_{i=1}^n a_i^{(k_i)}  - \prod_{i=1}^n a_i^{(k_i)} \Biggr] = 0, \nonumber
\end{align}
i.e., Eq.~\eqref{eqn:relation} is satisfied.  The functions $c_i(\eps)$ and $d_i(\eps)$ defined through Eq.~\eqref{eqn:Fk} thus satisfy $c_i(0) = 1$ and $d_i(0) = \overline{\alpha}_i$.

\subsubsection[3. Characterizing descending paths on stability landscape]{Characterizing descending paths on $\boldsymbol{\lambda^{\max}}$-landscape}
\label{si-subsec-C}

Here, we will apply the IFT to show that the functions $c_i(\eps)$ and $d_i(\eps)$ are continuously differentiable for small $\eps$ (and thus define a smooth curve in the neighborhood of the point $(1, \overline{\alpha}_i)$ in the $(c_i,d_i)$-plane), and we determine their first derivatives.  The condition under which we can apply the IFT to Eq.~\eqref{eqn:relation} at the point $(c_1,\ldots,c_n,d_1,\ldots,d_n,\eps) = (1,\ldots,1,\overline{\alpha}_1,\ldots,\overline{\alpha}_n,0)$ is that the $(2n-1) \times (2n-1)$ matrix
\begin{equation}\label{eqn:G}
\mathbf{G} \equiv 
\begin{pmatrix}
\frac{\partial F_1}{\partial c_1} & \cdots & \frac{\partial F_1}{\partial c_n} & \frac{\partial F_1}{\partial d_2} & \cdots & \frac{\partial F_1}{\partial d_n}\\
\vdots & & \vdots & \vdots & & \vdots\\
\frac{\partial F_{2n-1}}{\partial c_1} & \cdots & \frac{\partial F_{2n-1}}{\partial c_n} & \frac{\partial F_{2n-1}}{\partial d_2} & \cdots & \frac{\partial F_{2n-1}}{\partial d_n}\\
\end{pmatrix},
\end{equation}
is non-singular, where the elements of $\mathbf{G}$ are all evaluated at that point.  We note that $d_1$ is excluded from the set of variables here because $d_1 = 0$ always holds.  We also note that $\mathbf{G}$ (and whether it is singular or not) is completely determined by $\overline{\alpha}_2,\ldots,\overline{\alpha}_n$, and hence by the matrix $\mathbf{P}$ (see examples in Sec.~\ref{si-subsec-D}).  For notational convenience, define $x_s = c_s$ for $s=1,\ldots,n$ and $x_s = d_{s-n+1}$ for $s = n+1,\ldots,2n-1$.  Differentiating Eq.~\eqref{eqn:Fk}, we find 
an expression for the $(k,s)$-element of $\mathbf{G}$:
\begin{multline}\label{eqn:Gks}
G_{ks} = \frac{\partial F_k}{\partial x_s}(1,\ldots,1,\overline{\alpha}_1,\ldots,\overline{\alpha}_n,0) \\
= \begin{cases}
\displaystyle\rule{0pt}{22pt} \sum_{\{k_i\}} \prod_{i \neq s} a_i^{(k_i)} \cdot \chi \bigl( s,k,\{k_i\} \bigr), & \text{if $s=1,\ldots,n$},\\
\displaystyle\rule{0pt}{22pt} \sum_{\{k_i\}} \prod_{i \neq \hat{s}} a_i^{(k_i)} \cdot \chi \bigl( s,k,\{k_i\} \bigr), & \text{if $s=n+1,\ldots,2n-1$},
\end{cases}
\end{multline}
where the summation is defined as in Eq.~\eqref{eqn:k-coeff}; we denote $\hat{s} \equiv s-n+1$; the values of $a_i^{(k_i)}$ are given by Eq.~\eqref{eqn:beta-tilde}; and we have defined
\begin{equation}\label{eqn:chi}
\chi \bigl( s,k,\{k_i\} \bigr) \equiv \begin{cases}
1 &\text{if $\sum_i k_i = k$, $k_s = 1$, and $s=1,\ldots,n$},\\
1 &\text{if $\sum_i k_i = k$, $k_{\hat{s}} = 0$, and $s=n+1,\ldots,2n-1$},\\
0 &\text{otherwise}.
\end{cases}
\end{equation}

Since the eigenvalues of $\mathbf{P}$ are assumed to be distinct, we have $\frac{1}{4}=\overline{\alpha}_2<\cdots<\overline{\alpha}_n$.  We have numerically verified that this condition holds for all networks considered in the main text.  We first seek to show that the matrix $\mathbf{G}$ is non-singular by proving that the $s$th and $s'$th columns of $\mathbf{G}$ are linearly independent for all distinct pairs $s$ and $s'$.  To do this, we first note that Eq.~\eqref{eqn:Gks} simplifies in the special cases $k=1$, $k=2$, $k=2n-2$, and $k=2n-1$, as follows.

For $k=1$, to satisfy $\sum_i k_i = k$, we must have $k_i = 1$ for exactly one value of $i$ and $k_i = 0$ for all the others (recall that each $k_i$ is either $0$, $1$, or $2$).  For the case $1 \le s \le n$, we can derive a simplified formula, but it is not needed below, so we will skip that case here.  For the case $n+1 \le s \le 2n-1$, to have $\chi \bigl( s,k,\{k_i\} \bigr)=1$ in Eq.~\eqref{eqn:chi} we must have $k_t=1$ for some $t \neq \hat{s}$ and $k_i=0$ for all $i \neq t$ (including $i=\hat{s}$).  Since there are $n-1$ possibilities for $t$, there are that many nonzero terms in the summation in Eq.~\eqref{eqn:Gks}, which reduces to
\begin{equation}\label{eqn:G1s}
G_{1s} = \sum_{t\neq \hat{s}} \prod_{i\neq \hat{s}} a_i^{(k_i)} = \sum_{t\neq \hat{s}} \prod_{i\neq t,\hat{s}} \overline{\alpha}_i,
\quad n+1 \le s \le 2n-1.
\end{equation}

For $k=2$, the condition $\sum_i k_i = k$ implies that we either have (a) $k_i = 2$ for exactly one value of $i$ and $k_i = 0$ for all the others, or (b) $k_i = 1$ for two different values of $i$, and $k_i = 0$ for all the others.  For $1 \le s \le n$, we have $\chi \bigl( s,k,\{k_i\} \bigr) = 0$ for the terms in Eq.~\eqref{eqn:Gks} corresponding to case~(a), according to Eq.~\eqref{eqn:chi}.  For nonzero terms in Eq.~\eqref{eqn:Gks} corresponding to case~(b), we have $k_s=1$, $k_t=1$ with some $t\neq s$, and $k_i = 0$ for all $i \neq s,t$.  Putting the two cases together, we obtain 
\begin{equation}\label{eqn:G2s}
G_{2s} = \sum_{t\neq s} \prod_{i\neq s} a_i^{(k_i)} = \sum_{t\neq s} \prod_{i\neq t,s} \overline{\alpha}_i,
\quad 1 \le s \le n.
\end{equation}
A simplified expression for $G_{2s}$ for $n+1 \le s \le 2n-1$ can also be obtained, but it is not needed for our purpose.

For $k=2n-2$, satisfying $\sum_i k_i = k$ requires that either (a) $k_i = 0$ for exactly one value of $i$ and $k_i = 2$ for all the others, or (b) $k_i = 1$ for two different values of $i$, and $k_i = 2$ for all the others.  For $1 \le s \le n$, similarly to the case of $k=2$, there is no nonzero term in Eq.~\eqref{eqn:Gks} corresponding to case~(a), and for the nonzero terms in Eq.~\eqref{eqn:Gks} corresponding to case~(b), we have $k_s=1$, $k_t=1$ with some $t\neq s$, and $k_i = 2$ for all $i \neq s,t$.  Putting these two cases together, we obtain 
\begin{equation}\label{eqn:G2n-2s}
G_{2n-2,s} = \sum_{t\neq s} \prod_{i\neq s} a_i^{(k_i)} = \sum_{t\neq s} 1 = n-1,
\quad 1 \le s \le n.
\end{equation}
For $n+1 \le s \le 2n-1$, case (b) does not correspond to any nonzero term in Eq.~\eqref{eqn:Gks} because $k_i \neq 0$ for all $i$.  Case (a), on the other hand, yields exactly one term with $k_{\hat{s}} = 0$, leading to
\begin{equation}
G_{2n-2,s} = \prod_{i\neq \hat{s}} a_i^{(k_i)} = 1,
\quad n+1 \le s \le 2n-1.
\end{equation}

For $k=2n-1$, the only way to satisfy $\sum_i k_i = k$ is to have just one $k_i = 1$ and all the other $k_i = 2$.  For $1 \le s \le n$, we must have $k_s = 1$ for a nonzero term, so only one term survives in Eq.~\eqref{eqn:Gks}, and hence
\begin{equation}\label{eqn:G2n-1s}
G_{2n-1,s} = \prod_{i\neq s} a_i^{(k_i)} = 1,
\quad 1 \le s \le n.
\end{equation}
For $n+1 \le s \le 2n-1$, there is no nonzero term since there is no $i$ for which $k_i=0$ (see Eq.~\eqref{eqn:chi}), and thus
\begin{equation}\label{eqn:G2n-1s2}
G_{2n-1,s} = 0,
\quad n+1 \le s \le 2n-1.
\end{equation}

Now we can use Eqs.~\eqref{eqn:G1s}--\eqref{eqn:G2n-1s2} to show that any pair of columns of $\mathbf{G}$ are linearly independent.  For $1\le s \le n$ and $n+1 \le s' \le 2n-1$, the last two components ($k=2n-2$ and $k=2n-1$) of the $s$th and $s'$th column vectors form the two-dimensional vectors $(n-1, 1)^T$ and $(1,0)^T$, respectively, which are linearly independent. This implies that the full $(2n-1)$-dimensional vectors in the $s$th and $s'$th columns of $\mathbf{G}$ are also linearly independent.  For $1\le s < s' \le n$, since the last components ($k=2n-1$) of the $s$th and $s'$th column vectors are both equal to one, it suffices to show that the second component ($k=2$) is different in order to establish that they are linearly independent.  From Eq.~\eqref{eqn:G2s}, we have
\begin{equation}
\begin{split}
G_{2s} - G_{2s'} &= 
\sum_{t\neq s} \prod_{i\neq t,s} \overline{\alpha}_i
- \sum_{t\neq s'} \prod_{i\neq t,s'} \overline{\alpha}_i\\
&= \Bigl(\sum_{t\neq s,s'} \prod_{i\neq t,s} \overline{\alpha}_i + \prod_{i\neq s',s} \overline{\alpha}_i\Bigr)
- \Bigl(\sum_{t\neq s,s'} \prod_{i\neq t,s'} \overline{\alpha}_i + \prod_{i\neq s,s'} \overline{\alpha}_i\Bigr)\\
&= \sum_{t\neq s,s'} \Bigl(\prod_{i\neq t,s,s'} \overline{\alpha}_i \Bigr)(\overline{\alpha}_{s'} - \overline{\alpha}_s) > 0,
\end{split}
\end{equation}
since $\overline{\alpha}_{s'} > \overline{\alpha}_s$ and $\overline{\alpha}_i > 0$, $\forall i$, and hence all terms in the summation are positive. Thus, we have $G_{2s} \neq G_{2s'}$, implying that the $s$th and $s'$th column vectors are linearly independent.  For $n+1 \le s < s' \le 2n-1$, the argument is similar to the case of $1\le s < s' \le n$; it suffices to show that the first component ($k=1$) is different, and Eq.~\eqref{eqn:G1s} yields
\begin{equation}
\begin{split}
G_{1s} - G_{1s'} &= \sum_{t\neq \hat{s}} \prod_{i\neq t,\hat{s}} \overline{\alpha}_i - \sum_{t\neq \hat{s}'} \prod_{i\neq t,\hat{s}'} \overline{\alpha}_i\\
&= \sum_{t\neq \hat{s},\hat{s}'} \Bigl(\prod_{i\neq t,\hat{s},\hat{s}'} \overline{\alpha}_i \Bigr)(\overline{\alpha}_{\hat{s}'} - \overline{\alpha}_{\hat{s}}) > 0,
\end{split}
\end{equation}
where we have defined $\hat{s}' \equiv s'-n+1$ (and recall that $\hat{s} = s-n+1$).
Combining all of the above, we have that the $s$th and $s'$th column vectors of $\mathbf{G}$ are linearly independent for all pairs of distinct $s$ and $s'$, which establishes that $\mathbf{G}$ is non-singular.

The IFT can now be applied to Eq.~\eqref{eqn:relation} to conclude that the functions $c_i(\eps)$ and $d_i(\eps)$ are continuously differentiable.  Furthermore, their derivatives satisfy the set of equations obtained by substituting these functions into Eq.~\eqref{eqn:relation} and differentiating both sides with respect to $\eps$:
\begin{equation}
\frac{d}{d \eps} F_k(c_1(\eps),\ldots,c_n(\eps),d_1(\eps),\ldots,d_n(\eps),\eps) = 0.
\end{equation}
Using Eq.~\eqref{eqn:Fk}, this can be written as
\begin{multline}
\sum_{\{k_i\}} \chi \bigl( {\textstyle\sum_i k_i = k} \bigr)
\cdot \sum_{\ell=1}^n \Biggl[ \biggl(\prod_{i\neq\ell} a_i^{(k_i)}(\eps) \biggr) \cdot \frac{d a_\ell^{(k_\ell)}(\eps)}{d \eps} \,\,- \\
- \sum_{\sigma}\text{sign}(\sigma) \biggl( \prod_{i\neq\ell} E_{i\sigma(i)}^{(k_i)}(\eps) \biggr) \cdot \frac{d E_{\ell\sigma(\ell)}^{(k_\ell)}(\eps)}{d \eps} \Biggr] = 0,
\end{multline}
where we have now written the dependence of $a_i^{(k_i)}$ on $\eps$ explicitly.  Setting $\eps=0$ and using Eq.~\eqref{eqn:beta-tilde}, we obtain
\begin{equation}\label{eqn:deriv}
\sum_{\ell=1}^n \Biggl( \sum_{\{k_i\}} \prod_{i\neq\ell} a_i^{(k_i)}(0) \cdot \chi \bigl( {\textstyle\sum_i k_i = k} \bigr) \Biggr) \cdot \Biggl[ \frac{d a_\ell^{(k_\ell)}(0)}{d \eps}
- \frac{d E_{\ell\ell}^{(k_\ell)}(0)}{d \eps} \Biggr] = 0.
\end{equation}
Noting that $k_{\ell} = 0,1,2$ and that the derivatives are both zero if $k_{\ell} = 2$ (see Eqs.~\eqref{eqn:Eij} and \eqref{eqn:aik-def}), we can rearrange the summation in Eq.~\eqref{eqn:deriv} to write
\begin{multline}\label{eqn:deriv2}
\sum_{s=1}^n \Biggl( \sum_{\{k_i\}} \prod_{i \neq s} a_i^{(k_i)}(0) \cdot \chi \bigl( s,k,\{k_i\} \bigr) \Biggr) \cdot [x'_s(0) - y'_s(0)] \\
+ \sum_{s=n+1}^{2n-1} \Biggl( \sum_{\{k_i\}} \prod_{i \neq \hat{s}} a_i^{(k_i)}(0) \cdot \chi \bigl( s,k,\{k_i\} \bigr) \Biggr) \cdot [x'_s(0) - y'_s(0)] = 0,
\end{multline}
where we recall that $x_s(\eps) = c_s(\eps) = a_s^{(1)}(\eps)$ for $s=1,\ldots,n$ and $x_s(\eps) = d_{s-n+1}(\eps) = a_s^{(0)}(\eps)$ for $s = n+1,\ldots,2n-1$, and we use the notations $y_s(\eps) = \overline{B}_{ss}(\eps)$ for $s=1,\ldots,n$ and $y_s(\eps) = \overline{\alpha}_{s-n+1}$ for $s=n+1,\ldots,2n-1$.  
From Eq.~\eqref{eqn:Gks}, we see that Eq.~\eqref{eqn:deriv2} is equivalent to
\begin{equation}
\sum_{s=1}^{2n-1} G_{ks} [x'_s(0) - y'_s(0)] = 0,
\end{equation}
which can be put in vector form as $\mathbf{G}(\mathbf{x} - \mathbf{y}) = \mathbf{0}$ using the notations $\mathbf{x} = (x'_1(0),\dots, x'_{2n-1}(0))^T$ and $\mathbf{y} = (y'_1(0),\dots,y'_{2n-1}(0))^T$.  Thus, since $\mathbf{G}$ is non-singular, we have $\mathbf{x} = \mathbf{y}$, and hence $x'_i(0) = y'_i(0)$.  It then follows that
\begin{equation}
c'_i(0) = x'_i(0) = y'_i(0) = \frac{d \overline{B}_{ii}(0)}{d \eps} = \left.\frac{d}{d \eps} \left(\sum_{\ell=1}^n u_{i\ell} v_{i\ell} \gamma_\ell(\eps)/\beta_{=} \right) \right\vert_{\eps=0}
= \sum_{\ell=1}^n u_{i\ell} v_{i\ell} \gamma'_\ell(0)/\beta_{=}
\end{equation}
for $i=1,\ldots,n$, where we used the expression for $\overline{B}_{ij}$ from Eq.~\eqref{eqn:B-tilde}.  It also follows that
\begin{equation}
d'_i(0) = x'_{i+n-1}(0) = y'_{i+n-1}(0) = 
\left. \frac{d}{d \eps} \left( \overline{\alpha}_i \right) \right\vert_{\eps=0} = 0
\end{equation}
for $i=1,\ldots,n$, since $\overline{\alpha}_i$ is a constant that does not depend on $\eps$.  In particular, we have
\begin{equation}\label{eqn:c2e0-d2e0}
c'_2(0) = \sum_{\ell=1}^n u_{2\ell} v_{2\ell} \gamma'_\ell(0)/\beta_{=} \quad\text{and}\quad d'_2(0) = 0.
\end{equation}
Now, recall the argument at the end of Sec.~\ref{si-subsec-A} that $d'_2(0) = 0$ implies $c'_2(0) = 0$ if $\lambda^{\max}$ decreases along the path $\boldsymbol{\gamma}(\eps)$ in the $\boldsymbol{\upbeta}$-space.
In view of that argument, we see from Eq.~\eqref{eqn:c2e0-d2e0} that, if the eigenvalues of $\mathbf{P}$ are all distinct, the vector $(\gamma'_1(0),\dots, \gamma'_n(0))^T$ is parallel to the hyperplane $L$ whenever $\lambda^{\max}$ decreases along the path, where $L$ is uniquely defined by the equation
\begin{equation}\label{eqn:def_L}
\sum_{i=1}^n u_{2i} v_{2i} (\beta_i - \beta_{=}) = 0.
\end{equation}
In other words, any descending path on the $\lambda^{\max}$-landscape must be tangent to the hyperplane $L$ at $\boldsymbol{\upbeta}_{=}$.

If the path $\boldsymbol{\gamma}(\eps)$ is not tangent to $L$ (and hence $c'_2(0) \neq 0$), the expansion of $c_2(\eps)$ and $d_2(\eps)$ around $\eps=0$, which reads
\begin{equation}\label{eqn:c2_d2}
\begin{split}
c_2(\eps) &= 1 + c'_2(0)\eps + O(\eps^2),\\
d_2(\eps) &= \frac{1}{4} + O(\eps^2),
\end{split}
\end{equation}
can be substituted into Eq.~\eqref{eqn:lambda_max_eps} to obtain the following approximation for $\lambda^{\max}(\eps)$:
\begin{equation}\label{eqn:first_order}
\lambda^{\max}(\eps) =
\begin{cases}
\lambda^{\max}_{=} - \beta_{=} c'_2(0) \eps /2  + O(\eps^2), &\eps \le 0,\\ \lambda^{\max}_{=} + \beta_{=}\sqrt{c'_2(0) \eps /2\rule{0pt}{7.7pt}}   + O(\eps), &\eps > 0.
\end{cases}
\end{equation}
(Recall we are assuming $c'_2(0) \ge 0$.)  This establishes that the point $\boldsymbol{\upbeta}_{=}$ is a local minimizer of $\lambda^{\max}$ along any path that transversally intersects with the hyperplane $L$ at $\boldsymbol{\upbeta}_{=}$.

\subsubsection[4. Explicit calculation of matrix G for small n]{Explicit calculation of matrix $\mathbf{G}$ for small $\boldsymbol{n}$}
\label{si-subsec-D}

Here, we derive an expression for the matrix $\mathbf{G}$ in Eq.~\eqref{eqn:G} for $n = 2$ and for $n = 3$.  As in Sec.~\ref{si-subsec-A}--\ref{si-subsec-C} above,
we assume $\beta_{=} > 0$, or equivalently, $\alpha_2 > 0$.

\smallskip\noindent
{\bf Example 1}: 
For $n=2$, Eq.~\eqref{eqn:ch_poly} becomes
\begin{align}
\det(\overline{\mathbf{J}} - \nu\mathbf{I}) &= \det(\nu^2\mathbf{I} + \nu\overline{\mathbf{B}} + \mathbf{D}) = 
\det\begin{pmatrix}
\nu^2 + \overline{B}_{11}\nu & \overline{B}_{12}\nu \nonumber \\
\overline{B}_{21}\nu & \nu^2 + \overline{B}_{22}\nu + 1/4
\end{pmatrix}\\
&= (\nu^2 + \overline{B}_{11}\nu)(\nu^2 + \overline{B}_{22}\nu + 1/4) - \overline{B}_{12}\overline{B}_{21}\nu^2  \nonumber\\
&= \nu^4 + (\overline{B}_{11} + \overline{B}_{22})\nu^3
+ (\overline{B}_{11} \overline{B}_{22} - \overline{B}_{12}\overline{B}_{21} + 1/4) \nu^2
+ \overline{B}_{11} \nu/4,  \nonumber\\
&= \nu^4 + \frac{\beta_1 + \beta_2}{\beta_{=}}\,\nu^3
+ \left(\frac{\beta_1 \beta_2}{\beta_{=}^2} + \frac{1}{4}\right) \nu^2
+ \overline{B}_{11} \nu/4, \nonumber
\end{align}
while Eq.~\eqref{char_poly} becomes
\begin{equation}
\begin{split}
\det(\overline{\mathbf{J}} - \nu\mathbf{I}) &= 
(\nu^2 + c_1 \nu)(\nu^2 + c_2 \nu + d_2) \\
&= \nu^4 + (c_1 + c_2)\nu^3 + (c_1 c_2 + d_2)\nu^2 + c_1 d_2 \nu.
\end{split}
\end{equation}
Thus, by comparing coefficients for the same powers of $\nu$, we have
\begin{equation}\label{eqn:relation2}
\begin{split}
F_1(c_1,c_2,d_2,\eps) &= c_1 d_2 - \overline{B}_{11}/4,\\
F_2(c_1,c_2,d_2,\eps) &= c_1 c_2 + d_2 - \left(\frac{\beta_1 \beta_2}{\beta_{=}^2} + \frac{1}{4}\right),\\
F_3(c_1,c_2,d_2,\eps) &= c_1 + c_2 - \frac{\beta_1 + \beta_2}{\beta_{=}},  \nonumber
\end{split}
\end{equation}
and
\begin{gather*}
\frac{\partial F_1}{\partial c_1} = d_2, \quad \frac{\partial F_1}{\partial c_2} = 0, \quad
\frac{\partial F_1}{\partial d_2} = c_1,\\
\frac{\partial F_2}{\partial c_1} = c_2, \quad \frac{\partial F_2}{\partial c_2} = c_1, \quad
\frac{\partial F_2}{\partial d_2} = 1,\\
\frac{\partial F_3}{\partial c_1} = 1, \quad \frac{\partial F_3}{\partial c_2} = 1, \quad
\frac{\partial F_3}{\partial d_2} = 0,
\end{gather*}
\begin{equation}
\frac{\partial \mathbf{F}}{\partial (c,d)} =\begin{pmatrix}
d_2 & 0 & c_1\\
c_2 & c_1 & 1\\
1 & 1 & 0
\end{pmatrix}, \quad
\mathbf{G} = \frac{\partial \mathbf{F}}{\partial (c,d)} \biggr\vert_{\eps=0} =\begin{pmatrix}
1/4 & 0 & 1\\
1 & 1 & 1\\
1 & 1 & 0
\end{pmatrix}.  \nonumber
\end{equation}
Therefore, $\mathbf{G}$ is non-singular (regardless of the value of $\alpha_2>0$).

\medskip\noindent
{\bf Example 2}: 
For $n=3$, it is convenient to first write out the terms in the first sum of the first line in Eq.~\eqref{eqn:Fk} and differentiate them with respect to $c_i$ and $d_i$ (since the second sum does not contain $c_i$ or $d_i$). The terms from the first sum are
\begin{align}
F_1 &= c_1 d_2 d_3 - \cdots,\\
F_2 &= d_2 d_3 + c_1 c_2 d_3 + c_1 d_2 c_3 - \cdots,\\
F_3 &= c_1 c_2 c_3 + c_1 d_3 + c_1 d_2 + c_2 d_3 + d_2 c_3 - \cdots,\\
F_4 &= d_2 + d_3 + c_1 c_2 + c_2 c_3 + c_3 c_1 - \cdots,\\
F_5 &= c_1 + c_2 + c_3 - \cdots.
\end{align}
Using this, we see that the matrix of partial derivatives is
\begin{equation}
\frac{\partial \mathbf{F}}{\partial (c,d)} = \begin{pmatrix}
d_2 d_3 & 0 & 0 & c_1 d_3 & c_1 d_2\\
c_2 d_3 + d_2 c_3 & c_1 d_3 & c_1 d_2 & d_3 + c_1 c_3 & d_2 + c_1 c_2\\
c_2 c_3 + d_2 + d_3 & c_1 c_3 + d_3 & c_1 c_2 + d_2 & c_1 + c_3 & c_1 + c_2\\
c_2 + c_3 & c_3 + c_1 & c_1 + c_2 & 1 & 1\\
1 & 1 & 1 & 0 & 0
\end{pmatrix},
\end{equation}
from which we obtain the matrix $\mathbf{G}$ by setting $c_1 = c_2 = c_3 = 1$, $d_2 = \overline{\alpha}_2 = 1/4$, and $d_3 = \overline{\alpha}_3$:
\begin{equation}
\mathbf{G} = \begin{pmatrix}
\overline{\alpha}_3/4 & 0 & 0 & \overline{\alpha}_3 & 1/4\\
1/4 + \overline{\alpha}_3 & \overline{\alpha}_3 & 1/4 & 1+\overline{\alpha}_3 & 5/4\\
5/4 + \overline{\alpha}_3 & 1 + \overline{\alpha}_3 & 5/4 & 2 & 2\\
2 & 2 & 2 & 1 & 1\\
1 & 1 & 1 & 0 & 0
\end{pmatrix}.
\end{equation}
From this, we obtain 
$\text{det}(\mathbf{G}) = - \overline{\alpha}_3 (1-4\overline{\alpha}_3)^2/64
= - \alpha_3 (\alpha_2 - \alpha_3)^2 / (256 \alpha_2^3)$ and see that $\mathbf{G}$ is non-singular if and only if $0<\alpha_2 < \alpha_3$.  This condition is indeed satisfied by the 3-generator system considered in the main text ($0 < \alpha_2 \approx 75.5 < \alpha_3 \approx 178.5$).

\end{document}